\documentclass[journal]{IEEEtran}

\usepackage{booktabs}
\usepackage{threeparttable}
\usepackage{amsmath,graphicx,cite,amssymb}
\usepackage{amsfonts,multirow,bm,array,setspace,stfloats}

\usepackage{graphicx,float,cite,amssymb}
\usepackage{graphics}\DeclareGraphicsExtensions{.pdf,.jpeg,.png,.jpg}   
\usepackage{multirow,bm,bbm,array,setspace,dsfont}
\usepackage{textcomp}
\usepackage{psfrag}
\usepackage{color}
\usepackage{pstricks,enumerate}
\usepackage{bbm,subfigure}
\usepackage{algorithm,algpseudocode}
\makeatletter
\newcommand{\distas}[1]{\mathbin{\overset{#1}{\kern\z@\sim}}}%
\newsavebox{\mybox}\newsavebox{\mysim}
\newcommand{\distras}[1]{%
  \savebox{\mybox}{\hbox{\kern1pt$\scriptstyle#1$\kern1pt}}%
  \savebox{\mysim}{\hbox{$\sim$}}%
  \mathbin{\overset{#1}{\kern\z@\resizebox{\wd\mybox}{\ht\mysim}{$\sim$}}}%
}
\makeatother
\ifCLASSINFOpdf
\else
\fi

\newtheorem{proposition}{Proposition}

\newtheorem{lemma}{Lemma}
\newtheorem{theorem}{Theorem}

\newtheorem{example}{Example}

\newtheorem{remark}{Remark}

\newcommand{\ba}{\bm a}

\newcommand{\bb}{\bm b}

\allowdisplaybreaks[4]

\begin{document}
%
\title{Coordinating Multiple Intelligent Reflecting Surfaces without Channel Information}
\author{
	\IEEEauthorblockN{
	Fan Xu, \IEEEmembership{Member,~IEEE}, Jiawei Yao, \IEEEmembership{Student Member,~IEEE}, Wenhai Lai, \IEEEmembership{Student Member,~IEEE},\\
    Kaiming Shen, \IEEEmembership{Member,~IEEE},
	Xin Li, Xin Chen, and Zhi-Quan Luo, \IEEEmembership{Fellow,~IEEE}
} 
\thanks{

Manuscript published in IEEE Transactions on Signal Processing, vol. 12, pp. 31--46, Jan. 2024. This work was presented in part at the 2023 IEEE International Conference on Communications (ICC) and in part at the 2023 IEEE International Workshop on Signal Processing Advances in Wireless Communications (SPAWC). \emph{(Corresponding author: Kaiming Shen.)}

Fan Xu is with Peng Cheng Laboratory, Shenzhen (e-mail: xuf02@pcl.ac.cn).

Jiawei Yao, Wenhai Lai, and Kaiming Shen are with the School of Science and Engineering, The Chinese University of Hong Kong, Shenzhen (e-mail: jiaweiyao@link.cuhk.edu.cn; wenhailai@link.cuhk.edu.cn; shenkaiming@cuhk.edu.cn).

Xin Li and Xin Chen are with Huawei Technologies (e-mail: razor.lixin@huawei.com; chenxin@huawei.com).

Zhi-Quan Luo is with The Chinese University of Hong Kong, Shenzhen, with Shenzhen Research Institute of Big Data, China, and with Peng Cheng Laboratory, Shenzhen (e-mail: luozq@cuhk.edu.cn).
}
}

%


\maketitle

\begin{abstract}
Conventional beamforming methods for intelligent reflecting surfaces (IRSs) or reconfigurable intelligent surfaces (RISs) typically entail the full channel state information (CSI). However, the computational cost of channel acquisition soars exponentially with the number of IRSs. To bypass this difficulty, we propose a novel strategy called blind beamforming that coordinates multiple IRSs by means of statistics without knowing CSI. Blind beamforming only requires measuring the received signal power at the user terminal for a sequence of randomly generated phase shifts across all IRSs. The main idea is to extract the key statistical quantity for beamforming by exploring only a small portion of the whole solution space of phase shifts. We show that blind beamforming guarantees a signal-to-noise ratio (SNR) boost of $\Theta(N^{2L})$ under certain conditions, where $L$ is the number of IRSs and $N$ is the number of reflecting elements per IRS. The proposed conditions for achieving the optimal SNR boost of $\Theta(N^{4})$ in a double-IRS system are much easier to satisfy than the existing ones in the literature. Most importantly, the proposed conditions can be extended to a fully general $L$-IRS system. The above result significantly improves upon the state of the art in the area of multi-IRS-assisted communication. Moreover, blind beamforming is justified via field tests and simulations. In particular, as shown in our field tests at 2.6 GHz, our method yields up to 17 dB SNR boost; to the best of our knowledge, this is the first time that the use of multiple IRSs gets verified in the real world.
\end{abstract}

\section{Introduction}\label{sec:overview}

\IEEEPARstart{I}{ntelligent} reflecting surface (IRS), aka reconfigurable intelligent surface (RIS), is an emerging wireless network device that aims to improve wireless environment by manipulating signal reflections \cite{Wu2019,Bjoernson2022,Xu2021, OurICC}. Owing to its much lower cost and much lower energy consumption, IRS can provide an alternative to small base-station and relay for enhancing throughput, coverage, connectivity, and reliability in future networks such as the industrial Internet of Things (IIoT). While the early studies \cite{Wu2019} concentrate on a single IRS, the current trend is towards the multi-IRS coordination \cite{zhang2019analysis}, \cite{mei2022intelligent}. Many existing methods in this field require the full channel state information (CSI), thus suffering the curse of dimensionality when IRSs are deployed extensively. To bypass this difficulty, we propose a novel strategy called \emph{blind beamforming} that is capable of optimizing phase shifts across multiple IRSs in the absence of CSI.

Our approach is inspired by the two recent works \cite{Arun_2020_RFocus}, \cite{blind_beamforming_twc}, which suggest the potential of optimizing phase shifts blindly for a single IRS without CSI. Given the whole solution space $\Omega$ of phase shifts (which is too large to explore fully), \cite{Arun_2020_RFocus}, \cite{blind_beamforming_twc} propose only testing a small subset of possible solutions $\mathcal S\subset \Omega$ at random, from which a statistical quantity (e.g., the conditional sample mean) of the received signal power can be obtained to help decide phase shifts. The resulting solution is not restricted to $\mathcal S$. While \cite{Arun_2020_RFocus}, \cite{blind_beamforming_twc} focus on a single IRS, this work aims at a full generalization of blind beamforming that accounts for multiple IRSs.

Because the number of channels is exponential in the number of IRSs, channel estimation is a tractable task only in some simple settings, e.g., when there are two IRSs \cite{channel_est_siso,channel_est_MIMO_MU_zbx,channel_est_MSE_min,channel_est_beamforming_ycs}, or when the multi-hop reflected channels are all neglected \cite{Keykhosravi2021}. Some studies are devoted to the overhead reduction for channel estimation in IRS systems, e.g., the deep learning method \cite{Zhang2023self} and the two-timescale optimization \cite{zhao2021two}. Aside from the computational difficulty, channel estimation for IRS also imposes a huge practical challenge because of the communication chip issue as well as the network protocol issue 
\cite{blind_beamforming_twc}. To the best of our knowledge, the existing prototype realizations of IRS \cite{Arun_2020_RFocus,pei2021ris,tran2020demonstration,kitayama2021research, Staat_2022_IRShield, chen2020active} seldom involve channel estimation. 

Actually, even if the exact CSI has been provided, it is still quite difficult to decide phase shifts for multiple IRSs. The difficulty arises from the fact that every multi-hop reflected channel is incident to more than one reflecting element (RE) of distinct IRSs and hence their phase shifts must be optimized jointly. To render the problem tractable, a common compromise \cite{Lu2022,Sun2021,Song2022,Yang2022,Xie2022,Cao2021,Karim2022,Li2020,Ning2022,Wei2022,Huang2022,Esmaeilbeig2022,multi_IRS_radar_comm,Wei2022a,Li2022,Asim2022,Ni2022} is to ignore the multi-hop channels. 
Many existing analyses and methods build upon this approximation, ranging from delay alignment \cite{Lu2022} to ergodic rate \cite{Sun2021}, secure transmission \cite{Song2022}, spectral efficiency \cite{Yang2022}, outage probability \cite{Xie2022,Cao2021}, and full-duplex transmission \cite{Karim2022}. The above approximation has also been extended to the multiple-user case for a variety of system design problems related to IRS, e.g., the sum rates maximization \cite{Li2020,Ning2022,Wei2022}, the IRS placement optimization \cite{Huang2022}, the target sensing \cite{Esmaeilbeig2022}, the joint sensing and communication \cite{multi_IRS_radar_comm,Wei2022a}, the joint unmanned aerial vehicles (UAV) and IRS aided transmission \cite{Li2022,Asim2022}, and the federated learning \cite{Ni2022}.

However, the above simplified channel model with multiple IRSs could be fundamentally flawed. If each signal reflection is incident to only one IRS, then the multiple IRSs distributed at the different positions can be basically thought of as a single IRS. As a result, the signal-to-noise ratio (SNR) boost is at most $\Theta(L^2N^2)$ according to \cite{blind_beamforming_twc}, where $L$ is the number of IRSs and $N$ is the number of REs of each IRS.
In contrast, this work shows that a much higher boost of $\Theta(N^{2L})$ can be reached by harnessing the multi-hop reflections. Actually, the previous work \cite{Han_double_IRS_beamforming_power_scaling} already shows that the two-hop channels play a crucial role in enabling an SNR boost of $\Theta(N^4)$ for a double-IRS system. Nevertheless, the argument in \cite{Han_double_IRS_beamforming_power_scaling} is based on a fairly strong assumption that only the two-hop reflections exist while the rest channels are all null. Similarly, \cite{Multi_IRS_Huang,multi_IRS_Mei} only assume the existence of the longest cascaded channels (which are incident to every IRS) from transmitter to receiver in a general $L$-IRS system. A line of other works \cite{multi_IRS_WMMSE,3D_channel_model_double_IRS,Proactive_Eavesdropping,MIMO_MU_Zhengbeixiong,MISO_Chen,Nguyen2022,Kim2021} simplify the multi-IRS channel model in the opposite way. They only consider the one-hop and the two-hop reflections while neglecting all the higher-order reflections. Differing from all the above works, this paper does not require any channels to be zero. As a major result of this work, we show that the highest possible SNR boost of $\Theta(N^{2L})$ can be achieved by blind beamforming without making any zero approximations of the channels. 

The main contributions of this paper are summarized below:
\begin{enumerate}
\item We propose a blind beamforming method for the double-IRS system. It extends the existing blind beamforming algorithm by performing the CSM method \cite{blind_beamforming_twc} for the two IRSs sequentially. Although this extension is natural and simple, its performance analysis is by no means trivial. We show that the optimal SNR boost $\Theta(N^4)$ can be achieved under the following conditions C1 (when multi-hop channels can be decomposed), C2 (when the phase shift resolution is sufficiently large), and C3 (when the multi-hop channels are sufficiently strong). The proposed optimality conditions are much easier to satisfy than the existing ones in \cite{Han_double_IRS_beamforming_power_scaling,Han2022}.
\item We further extend blind beamforming to the fully general case with $L$ IRSs. The extension is still straightforward: we perform CSM \cite{blind_beamforming_twc} sequentially across the $L$ IRSs. Again, the difficulty lies in the performance analysis. The main result in this part of our work is to extend the optimality conditions C1--C3 to the multi-IRS case as D1--D3. To the best of our knowledge, this is the first known set of nontrivial optimality conditions for more than two IRSs.
\item Aside from simulations, we conduct field tests to demonstrate the proposed blind beamforming method with multiple IRSs at 2.6 GHz. The test results show that our method can outperform the benchmarks significantly for both indoor and outdoor environments. To the best of our knowledge, this is the first time that the use of multiple IRSs is verified in the real world.
\end{enumerate}

\renewcommand{\arraystretch}{1.0}
\begin{table}[t]
\renewcommand{\arraystretch}{1.3}
\small
\centering
\caption{\small List of Main Variables}
\begin{tabular}{|l|l|}
\hline
Symbol & Definition \\
\hline
$L$ & number of IRSs   \\
\hline
$N$ & number of REs of each IRS\\
\hline
$T$ & number of random samples for blind beamforming\\
\hline
$n_{\ell}$ & index of the $n$th RE of IRS $\ell$\\
\hline
$K_\ell$ & number of phase shift choices on RE $n_\ell$\\
\hline
$h_{n_{1},\ldots,n_{L}}$ & cascaded channel induced by REs $n_1,\ldots,n_L$\\
\hline
$u^{(\ell)}_{n_\ell}$ & factor component of $h_{n_{1},\ldots,n_{L}}$ related to RE $n_\ell$\\
\hline
$\theta_{n_{\ell}}$ & phase shift of RE $n_{\ell}$\\
\hline
$\theta'_{n_{\ell}}$ & solution of $\theta_{n_{\ell}}$ by the proposed method\\
\hline
$\theta^\star_{n_{\ell}}$ & continuous solution of $\theta_{n_{\ell}}$ as $K\rightarrow\infty$\\
\hline
$\hat\theta^\star_{n_{\ell}}$ & approximate continuous solution of $\theta_{n_{\ell}}$\\
\hline
$\mathcal D^{(\ell)}_{m}$ & set of reflected channels related to RE $m$ of IRS $\ell$\\
\hline
$\mathcal D^{(\ell)}_{0}$ & set of channels not related to any RE of IRS $\ell$\\
\hline
$\mathcal A^{(\ell)}_{m}$ & subset of $\mathcal D^{(\ell)}_{m}$ unrelated to at least one IRS \\
\hline
$\mathcal E^{(\ell)}_{m}$ & subset of $\mathcal D^{(\ell)}_{m}$ related to every IRS \\
\hline
\end{tabular}
\label{tab:var}
\vspace{1em}
\end{table}

The remainder of the paper is organized as follows. Section \ref{sec:model} describes the multi-IRS channel model and formulates the beamforming problem mathematically. Section \ref{sec blind beamforming} introduces the blind beamforming method for a double-IRS system. Section \ref{sec csm} extends the proposed method to a general $L$-IRS system. Section \ref{sec test} shows field test and simulation. Finally, Section \ref{sec:conclusion} concludes this work. 

The Bachmann-Landau notation is extensively used in the paper: $f(n)=O(g(n))$ if there exists some $c>0$ such that $|f(n)|\le cg(n)$ for $n$ sufficiently large;
$f(n)=o(g(n))$ if for any $c>0$ it holds that $|f(n)|< cg(n)$ for $n$ sufficiently large; $f(n)=\Omega(g(n))$ if there exists some $c>0$ such that $f(n)\ge cg(n)$ for $n$ sufficiently large; $f(n)=\Theta(g(n))$ if $f(n)=O(g(n))$ and $f(n)=\Omega(g(n))$ both hold. Moreover, the phase of a complex number $x\in\mathbb C$ is written as $\angle x$, and the discrete set $\{a,a+1,\ldots,b-1,b\}$ is written as $[a:b]$ for two integers $a<b$.  The cardinality of a set $\mathcal{A}$ is written as $|\mathcal{A}|$; the absolute value of a complex number $a$ is written as $|a|$.  For convenience, we summarize in TABLE \ref{tab:var} the main variables used in the sequel.

\section{System Model}
\label{sec:model}
\begin{figure}[t]
\centering
\includegraphics[width=6.5cm]{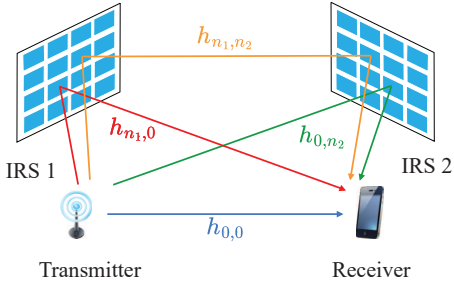}
\caption{A double-IRS system with $L=2$, where $h_{0,0}$ is the direct channel, $\{h_{n_1,0},h_{0,n_2}\}$ are the one-hop reflected channels, and $\{h_{n_1,n_2}\}$ are the two-hop reflected channels.}
\label{fig:system model}
\end{figure}

Consider a point-to-point wireless transmission in aid of $L\ge2$ IRSs. Assume that the transmitter and receiver are equipped with one antenna each. Assume also that every\footnote{We assume that IRSs have the same number of REs in order to facilitate performance analysis. But this assumption is not required for the practical use of blind beamforming as discussed in Section \ref{subsec:field_tests}.} IRS consists of $N$ REs. We use $\ell\in[1:L]$ to index each IRS, and use $n_\ell\in[1:N]$ to index each RE of IRS $\ell$. Let $\theta_{n_\ell}\in[0,2\pi)$ be the phase shift induced by RE $n_\ell$ into its associated reflected channels. From a practical stand \cite{blind_beamforming_twc,pei2021ris,tran2020demonstration,kitayama2021research, Staat_2022_IRShield, chen2020active}, assume that each $\theta_{n_\ell}$ of IRS $\ell$ can only take on values from a uniform discrete set
\begin{equation}
    \Phi_{K_\ell} = \{\omega,2\omega,\ldots,K_\ell\omega\}\;\; \text{where} \;\; \omega = \frac{2\pi}{K_\ell}
\end{equation}
given a positive integer $K_\ell\ge2$, namely \emph{discrete beamforming}. We use $h_{n_1,\ldots,n_L}$ to denote the cascaded reflected channel induced by the REs $(n_1,n_2,\ldots,n_L)$; let $n_\ell=0$ if the channel is not related to IRS $\ell$. For instance, if $L=3$ and $N=10$, then $h_{2,0,6}$ represents a reflected channel incident to the 2nd RE of IRS 1 and the 6th RE of IRS 3, which is not related to any RE of IRS 2. In particular, $h_{0,\ldots,0}$ represents the direct channel from the transmitter to the receiver.
For the transmit signal $X\in\mathbb C$ and the complex Gaussian background noise $Z\sim\mathcal{CN}(0,\sigma^2)$, the received signal $Y\in\mathbb C$ is given by
\begin{equation}
\label{Y:K-IRS}
    Y = \sum_{(n_1,\ldots,n_L)\in[0:N]^L}h_{n_1,\ldots,n_L}e^{j\sum^L_{\ell=1}\theta_{n_\ell}}X+Z.
\end{equation}
For each $n_\ell=0$, we accordingly set $\theta_{n_\ell}=0$. When specialized to the double-IRS case with $L=2$, the above equation can be rewritten as
\begin{multline}
\label{Y: 2-IRS}
    Y = \underbrace{h_{0,0}X}_{\text{direct signal}} + \underbrace{\sum^{N}_{n_1=1}h_{n_1,0}e^{j\theta_{n_1}}X}_{\text{reflected signal due to IRS 1}}+\underbrace{\sum^{N}_{n_2=1}h_{0,n_2}e^{j\theta_{n_2}}X}_{\text{reflected signal due to IRS 2}}\\
    +\underbrace{\sum^{N}_{n_1=1}\sum^{N}_{n_2=1}h_{n_1,n_2}e^{j(\theta_{n_1}+\theta_{n_2})}X}_{\text{reflected signal due to both IRS 1 \& IRS 2}}+Z,
\end{multline}
as illustrated in Fig.~\ref{fig:system model}. In most of this work, we assume a general integer $L\ge2$. Section \ref{sec blind beamforming} focuses on the special case of $L=2$.

With the transmit power $P=\mathbb E[|X|^2]$, the received SNR is
\begin{equation}
    \mathrm{SNR}=
    \left|\sum_{(n_1,\ldots,n_L)\in[0:N]^L}h_{n_1,\ldots,n_L}e^{j\sum^L_{\ell=1}\theta_{n_\ell}}\right|^2\frac{P}{\sigma^2}.
\end{equation}
We wish to evaluate the performance gain brought by the IRSs. Toward this end, let us also compute the SNR without using any IRS as a benchmark, that is
\begin{equation} \label{eqn SNR}
    \mathrm{SNR}_0=\left|h_{0,\ldots,0}\right|^2\frac{P}{\sigma^2}.
\end{equation}
We seek the optimal set of phase shifts $\{\theta_{n_\ell}\}$ that maximizes the SNR boost, i.e.,
\begin{subequations}\label{eqn problem}
\begin{align}
\underset{ \{\theta_{n_\ell}\}}{\text{maximize}}\quad & \frac{\mathrm{SNR}}{\mathrm{SNR}_0} \label{eqn problem1}\\ 
\textrm{subject to}\quad &\theta_{n_\ell}\in\Phi_{K_\ell},\;\forall n_\ell.\label{eqn problem2}
\end{align}
\end{subequations}
The difficulties of the above problem are two-fold: (i) the variables are discrete; (ii) the channels $\{h_{n_1,\ldots,n_L}\}$ are unknown.

\section{Double-IRS Case}\label{sec blind beamforming}

The conventional paradigm for IRS beamforming comprises two stages: first estimate the cascaded channels $\{h_{n_1,\ldots,n_L}\}$ and then optimize the phase shifts $\{\theta_{n_\ell}\}$. But channel acquisition does not scale well with problem size because the number of channels grows exponentially with the number of IRSs. Alternatively, one may just estimate the channel matrix between every pair of IRSs and subsequently recover the cascaded channels $\{h_{n_1,\ldots,n_L}\}$ by multiplying the associated between-IRS channel matrices together, so that the number of channels to estimate decreases to $2NL+{L \choose 2}N^2=O(N^2L^2)$. However, the above method is costly in practice because it requires deploying a sensor at each RE to detect the pilot signal for channel estimation. Differing from most approaches in the literature, this work sidesteps channel estimation and optimizes phase shifts directly in the absence of CSI.

\subsection{Blind Beamforming for a Single IRS}

Before proceeding to the double-IRS case, we first review the so-called \emph{conditional sample mean (CSM)} method in \cite{blind_beamforming_twc} for configuring a single IRS without any channel information. We then let $L=1$. Since there is only one IRS, the IRS index $\ell$ can be dropped for ease of notation, i.e., $n_\ell$ reduces to $n$.

If all the channels were already known, then a natural idea would be to align each reflected channel $h_n$ with the direct channel $h_0$. If the perfect alignment cannot be achieved due to the discrete constraint $\Phi_{K}$, one may rotate $h_n$ to the closest possible position to $h_{0}$ in the complex plane, namely the \emph{closest point projection (CPP)}, whereby phase shift is determined as
\begin{equation}
\label{CPP}
    \theta^{\text{CPP}}_{n} = \arg\min_{\theta\in\Phi_{K}}\big|\theta+\angle h_n-\angle h_0\big|.
\end{equation}
The aim of CSM is to mimic CPP without knowing $\angle h_n$ and $\angle h_0$. The method works as follows. We first generate a total of $T$ random samples $\bm\theta^{(t)}=\{\theta^{(t)}_{n}\,\big|\,n\in[1:N]\}$ with each $\theta^{(t)}_{n}$ drawn uniformly from $\Phi_{K}$, for the sample index $t\in[1:T]$. Let $\mathcal G_{n,k}\subseteq[1:T]$ be the set of indices of those samples $\bm\theta^{(t)}$ satisfying $\theta^{(t)}_{n}=k\omega$, i.e.,
\begin{align}
    \mathcal{G}_{n,k}=\Big\{ t\in[1:T]\,\Big|\,\theta^{(t)}_{n}=k\omega\Big\}.
\end{align}
We measure the received signal power $|Y^{(t)}|^2$ corresponding to each random sample $\bm\theta^{(t)}$, based on which a conditional sample mean of $|Y^{(t)}|^2$ is computed for each $\mathcal G_{n,k}$ as
\begin{equation}
\label{cond_expectation}
    \widehat{\mathbb E}[|Y|^2\,|\,\theta_{n}=k\omega]=\frac{1}{|\mathcal{G}_{n,k}|} \sum\limits_{t \in \mathcal{G}_{n,k}} |Y^{(t)}|^2.
\end{equation}
The solution by CSM, denoted $\theta'_n$, maximizes the conditional sample mean with respect to each RE, i.e.,
\begin{align}
\label{CSM}
    \theta'_n = \arg \max_{\varphi\in\Phi_K} \widehat{\mathbb{E}}[|Y|^2\,|\,\theta_n=\varphi].
\end{align}
Define the average-reflection-to-direct-signal ratio to be
\begin{equation}
    \rho = \frac{\sum^N_{n=1}{|h_n|^2}/N}{|h_0|^2},
\end{equation}
which converges in probability to a finite constant for $N$ sufficiently large. The behavior of CSM is characterized by the following proposition.
\begin{proposition}[Theorem 2 in \cite{blind_beamforming_twc}]
\label{prop:CSM}
The CSM method is equivalent to the CPP method in \eqref{CPP} and yields a quadratic SNR boost in the number of REs in expectation, i.e.,
\begin{equation}
\label{boost:single_IRS}
    \mathbb E\left[\frac{\mathrm{SNR}}{\mathrm{SNR}_0}\right] = \rho\cdot\Theta(N^2),
\end{equation}
so long as $K\ge3$ and $T=\Omega(N^2(\log N)^3)$, where the expectation is taken over random samples of $(\theta_1,\ldots,\theta_N)$.
\end{proposition}

\begin{remark}
CSM only requires trying out a polynomial number of possible solutions $(\theta_1,\ldots,\theta_n)$, which occupy a small portion of the whole solution space of size $K^N$.
\end{remark}

\begin{remark}
For the binary beamforming case with $K=2$, i.e., when each $\theta_n\in\{0,\pi\}$, the SNR boost by CSM may fall below the quadratic. Rather, the SNR boost can be arbitrarily close to 0 dB in the worst-case scenario as shown in \cite{blind_beamforming_twc}. In contrast, an enhanced CSM in \cite{blind_beamforming_twc} maintains the quadratic boost for any $K\ge2$. Nevertheless, the contrived worst-case scenario of CSM rarely occurs in practice, so CSM is still a good choice for binary beamforming.
\end{remark}

\subsection{Blind Beamforming for Two IRSs}
\label{subsec:L=2}

We now let $L=2$. The above CSM method is extended to the double-IRS case as follows: first optimize IRS 1 while holding IRS 2 fixed, and then optimize IRS 2 while holding IRS 1 fixed, as summarized in Algorithm \ref{alg:SCSM_double}. Simple as the extension looks, it is by no means trivial to analyze its performance. The main result of this subsection is to establish a quartic SNR boost of $\Theta(N^4)$ under certain conditions.

\begin{algorithm}[t]
\caption{Blind Beamforming for a Double-IRS System}
\label{alg:SCSM_double}
\begin{algorithmic}[1]
\State{Initialize all the phase shifts to zero.}
\State{Generate $T$ random samples for IRS 1.}
\For{$t=1,\ldots,T$}
\State{Measure the received signal power $|Y^{(t)}|^2$.}
\EndFor
\For{$n_1=1,\ldots,N$}
\For{$k=1,\ldots,K_1$}
\State{Compute the conditional sample mean in \eqref{cond_expectation}.}
\EndFor
\State Decide each $\theta_{n_{1}}$ for IRS $1$ according to \eqref{CSM}.
\EndFor
\State{Generate $T$ random samples for IRS 2.}
\For{$t=1,\ldots,T$}
\State{Measure the received signal power $|Y^{(t)}|^2$.}
\EndFor
\For{$n_2=1,\ldots,N$}
\For{$k=1,\ldots,K_2$}
\State{Compute the conditional sample mean in \eqref{cond_expectation}.}
\EndFor
\State Decide each $\theta_{n_{2}}$ for IRS $2$ according to \eqref{CSM}.
\EndFor
\end{algorithmic}
\end{algorithm}

Let us start with a common misconception. One may think that the SNR boost can be readily verified for the extended CSM because each IRS brings a $\Theta(N^2)$ boost as shown in Proposition \ref{prop:CSM} and hence the two IRSs together bring a $\Theta(N^4)$ boost. The above argument is problematic in that the boost factor $\rho$ in \eqref{boost:single_IRS} associated with IRS 1 is impacted by the later optimization of IRS 2. As a quick example, if the channels between the two IRSs are all zeros so that only $h_{0,0},h_{n_1,0},h_{0,n_2}$ survive, then the two IRSs can be recognized as one whole IRS, and thus the highest possible boost is $\Theta(N^2)$. The reason is that each $h_{0,n_2}$ is included in the fixed direct channel when analyzing the SNR boost for IRS 1, but subsequently it can be altered dramatically by the optimization of IRS 2. Thus, the key question is: how do we preserve the SNR boost of the previous IRS when optimizing the current IRS? The following theorem provides a set of sufficient conditions in this regard.


\begin{theorem}
\label{coro boost 2}
If a double-IRS system satisfies the following three conditions:
\begin{enumerate}[C1.]
    \item the channels between the two IRSs are  
    line-of-sight (LoS) so that the two-hop channel matrix has rank one \cite{Han_double_IRS_beamforming_power_scaling} and can be factorized as 
    \begin{equation}
    \label{eqn 2 decompose}
        \begin{bmatrix}
    h_{1,1} & \cdots & h_{1,N}\\
    \vdots &  & \vdots\\
    h_{N,1} & \cdots & h_{N,N}
    \end{bmatrix}
    =
    \begin{bmatrix}
    u^{(1)}_1\\
    \vdots\\
    u^{(1)}_N
    \end{bmatrix}
    \begin{bmatrix}
    u^{(2)}_1 &\cdots &
    u^{(2)}_N
    \end{bmatrix},
    \end{equation}
    where $u_n^{(1)}\neq 0$ and $u_n^{(2)}\neq 0$ for all $n\in[1:N]$;
    \item $K_1,K_2\ge3$;
    \item there exists a constant $\gamma\in[0,\frac{\pi}{2}-\frac{\pi}{K_1})$ such that
    \begin{equation}
    \label{gamma}
    |h_{n_1,0}| \le \sin\gamma \cdot\left|\sum^N_{n_2=1} h_{n_1,n_2}\right|,\;\;\forall n_1\in[1:N],
    \end{equation}
\end{enumerate}
then the extended CSM method as stated at the beginning of this subsection yields a quartic SNR boost as
    \begin{align}
    \mathbb{E}\left[\frac{\mathrm{SNR}}{\mathrm{SNR}_0}\right]=\frac{\delta_1^2\delta_2^2}{|h_0|^2}\cdot\Theta(N^4)
    \label{eqn n4gain}
    \end{align}
so long as $T=\Omega(N^2(\log N)^3)$, where the expectation is taken over random samples of phase shifts, and
\begin{equation}
    \delta_1\triangleq\frac{1}{N}\sum^N_{n_1=1}|u^{(1)}_{n_1}|\quad\text{and}\quad
    \delta_2\triangleq\frac{1}{N}\sum^N_{n_2=1}|u^{(2)}_{n_2}|.
\end{equation}
\end{theorem}

\begin{IEEEproof}
Some concrete examples are provided later in Section \ref{subsec:prev_work} to illustrate the proposed conditions C1--C3. A formal proof follows. Since $|h_0|^2$	and $P$ are fixed, it suffices to show that $\mathbb E[|g|^2]=\delta^2_1\delta^2_2\cdot\Theta(N^4)$, where $g$ represents the superposition of all the channels with the IRS phase shifts $\theta_{n_1}$ and $\theta_{n_2}$, i.e.,
\begin{multline}
\label{g L=2}
g(\theta_{n_1},\theta_{n_2}) = h_{0,0}+\sum^N_{n_1=1}h_{n_1,0}e^{j\theta_{n_1}}+\sum^N_{n_2=1}h_{0,n_2}e^{j\theta_{n_2}}\\
+\sum^N_{n_1=1}\sum^N_{n_2=1}h_{n_1,n_2}e^{j(\theta_{n_1}+\theta_{n_2})}.
\end{multline}
To establish $\mathbb E[|g|^2]=\delta^2_1\delta^2_2\cdot\Theta(N^4)$, we need to verify the converse $\mathbb E[|g|^2]=\delta^2_1\delta^2_2\cdot O(N^4)$ and the achievability $\mathbb E[|g|^2]=\delta^2_1\delta^2_2\cdot \Omega(N^4)$. The converse is evident since
\begin{align}
|g|^2&\le\!\bigg||h_{0,0}|\!+\!\!\sum_{n_1=1}^N\!|h_{n_1,0}|\!+\!\!\sum_{n_2=1}^N\!|h_{0,n_2}|\!+\!\!\sum_{n_1=1}^N\sum_{n_2=1}^N\!|h_{n_1,n_2}|\bigg|^2\notag\\
&=\delta_1^2\delta_2^2\cdot O(N^4).\notag
\end{align}
The rest of the proof focuses on the achievability.

According to the extended CSM method stated at the beginning of this subsection, we first configure IRS 1 with the phase shifts of IRS 2 fixed to 0, by treating all the channels related to IRS 1 as the reflected channel and the rest as the direct channel.
Thus, if $\theta_{n_1}$ is continuous, its optimal solution is aligning the reflected channel with the direct channel exactly, i.e.,
\begin{equation}
    \theta^\star_{n_1} = \angle\Bigg(\underbrace{h_{0,0}+\sum_{n_2=1}^N h_{0,n_2}}_{\text{direct channel}}\Bigg)-\angle\Bigg(\underbrace{h_{n_1,0}+\sum_{n_2=1}^N h_{n_1,n_2}}_{\text{reflected channel}}\Bigg).\notag
\end{equation}
When $T=\Omega(N^2(\log N)^3)$, according to Proposition \ref{prop:CSM}, configuring IRS 1 by conditional sample mean is equivalent to rotating the reflected channel to the closest position to the direct channel, i.e.,
\begin{equation}
    \theta'_{n_1} = \arg\min_{\theta\in\Phi_{K_1}}\big|\theta-\theta^\star_{n_1}\big|.
\end{equation}
Clearly, we have
\begin{equation}
\label{bound:1}
    \big|\theta'_{n_1}-\theta^\star_{n_1}\big|\le\frac{\pi}{K_1}.
\end{equation}
Moreover, we approximate the continuous solution $\theta^\star_{n_1}$ by removing the single-hop reflect channels, i.e., 
\begin{align}
    \hat\theta^\star_{n_1} &=\angle\left(h_{0,0}+\sum_{n_2=1}^N h_{0,n_2}\right)-\angle\left(\sum_{n_2=1}^N h_{n_1,n_2}\right)\notag\\
&=\angle\Bigg(h_{0,0}+\sum_{n_2=1}^N h_{0,n_2}\Bigg)-\angle u^{(1)}_{n_1}-\angle\!\left(\sum_{n_2=1}^N u^{(2)}_{n_2}\right),\label{eqn 2 theta^}
\end{align}
where the second equality follows from the assumption in \eqref{eqn 2 decompose}. Next, we show that $\hat\theta^\star_{n_1}$ can approximate $\theta'_{n_1}$ closely so long as the conditions C2 and C3 both hold.

With the inequality in \eqref{gamma}, we can bound the difference between $\hat\theta^\star_{n_1}$ and $\theta^\star_{n_1}$ as
\begin{align}
    \big|\hat\theta^\star_{n_1}-\theta^\star_{n_1}\big|&=\left|\angle\left(h_{n_1,0}\!+\!\sum_{n_2=1}^N h_{n_1,n_2}\right)\!-\!\angle\left(\sum_{n_2=1}^N h_{n_1,n_2}\right)\right|\notag\\
&\le \gamma,\label{eqn 2 gamma}
\end{align}
where the second step is explained in Fig. \ref{fig:explanation L=2}. Now, with $K_1\ge3$ in C2 and $\gamma\in[0,\frac{\pi}{2}-\frac{\pi}{K_1})$ in C3, combining \eqref{bound:1} and \eqref{eqn 2 gamma} together gives
\begin{equation}
\label{bound:3}
    \big|\hat\theta^\star_{n_1}-\theta'_{n_1}\big|\le \gamma+\frac{\pi}{K_1}<\frac{\pi}{2}.
\end{equation}

\begin{figure}[t]
    \centering
    \includegraphics[width=7cm]{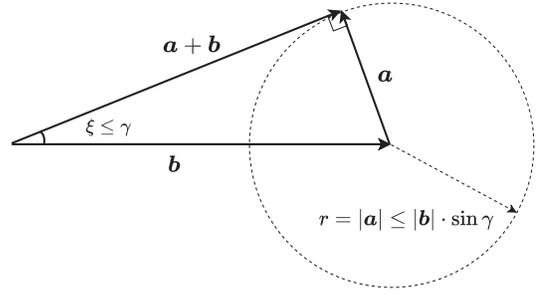}
    \caption{Illustration of the second step in \eqref{eqn 2 gamma}. Let $\ba=h_{n_1,0}$ and $\bb=\sum_{n_2=1}^N h_{n_1,n_2}$. According to \eqref{gamma}, $\ba$ must lie on a circle with its radius smaller than $(\sin\gamma)\cdot|\bb|$, so the angle between $\bb$ and $\ba+\bb$ is no greater than $\gamma$ as can be seen from the geometry.}
    \label{fig:explanation L=2}
\end{figure}

Next, IRS 2 is configured with each phase shift of IRS 1 fixed at $\theta'_{n_1}$. We now treat all the channels related to IRS 2 as reflected channel and treat the rest as the direct channel. Thus, if $\theta_{n_2}$ is continuous, its optimal solution is given by
\begin{align}
    &\theta^\star_{n_2} =\notag\\ &\;\angle\Bigg(\!\underbrace{h_{0,0}\!+\!\sum_{n_1=1}^N h_{n_1,0}e^{j\theta'_{n_1}}}_{\text{direct channel}}\!\Bigg)
    \!-\!\angle\Bigg(\!\underbrace{h_{0,n_2}\!+\!\sum_{n_1=1}^N h_{n_1,n_2}e^{j\theta'_{n_1}}}_{\text{reflected channel}}\!\Bigg).\label{eqn 2 }
\end{align}
Again, when $T=\Omega(N^2(\log N)^3)$,  Proposition \ref{prop:CSM} gives 
\begin{equation}
    \theta'_{n_2} = \arg\min_{\theta\in\Phi_{K_2}}\big|\theta-\theta^\star_{n_2}\big|
\end{equation}
and
\begin{equation}
\label{bound:4}
    \big|\theta'_{n_2}-\theta^\star_{n_2}\big|\le\frac{\pi}{K_2}.
\end{equation}
For ease of notation, we define
\begin{equation}
    \xi_{n_2} = h_{0,n_2}+\sum_{n_1=1}^N h_{n_1,n_2}e^{j\theta'_{n_1}}.\label{eqn xi}
\end{equation}
It can be shown that
\begin{align}
    |g(\theta'_1,\theta'_2)|^2 &= \left|h_{0,0}+\sum_{n_1=1}^N h_{n_1,0}e^{j\theta'_{n_1}}+\sum_{n_2=1}^{N}e^{j\theta'_{n_2}}\xi_{n_2}\right|^2\notag\\
    &\ge \left(\cos\frac{\pi}{K_2}\cdot\sum_{n_2=1}^N \left|\xi_{n_2}\right|\right)^2,
    \label{key_step:1}
\end{align}
where the last inequality follows by the projection of each $e^{j\theta'_{n_2}}\xi_{n_2}$ onto $h_{0,0}+\sum_{n_1=1}^N h_{n_1,0}e^{j\theta'_{n_1}}$ and by the fact that the angle between them is bounded above by $\pi/K_2$ according to \eqref{bound:4}. We further bound the $|\xi_{n_2}|$ as follows:
\begin{align}
    |\xi_{n_2}| &= \left|h_{0,n_2}+\sum_{n_1=1}^N h_{n_1,n_2}e^{j\theta'_{n_1}}\right|\notag\\
    &= \left|\sum_{n_1=1}^N h_{n_1,n_2}e^{j\theta'_{n_1}}\right|+o(N)\notag\\
    &= \left|\sum_{n_1=1}^N h_{n_1,n_2}e^{j\hat\theta^\star_{n_1}}e^{j(\theta'_{n_1}-\hat\theta^\star_{n_1})}\right|+o(N)\notag\\
    &\overset{(a)}{=} \left|\sum_{n_1=1}^N u^{(1)}_{n_1}u^{(2)}_{n_2}e^{j(\eta-\angle u^{(1)}_{n_1})}e^{j(\theta'_{n_1}-\hat\theta^\star_{n_1})}\right|+o(N)\notag\\
    &= |u^{(2)}_{n_2}|\cdot\left|\sum_{n_1=1}^N |u^{(1)}_{n_1}|e^{(\theta'_{n_1}-\hat\theta^\star_{n_1})}\right|+o(N)\notag\\
    &\overset{(b)}{\ge} |u^{(2)}_{n_2}|\cdot\cos\left(\gamma+\frac{\pi}{K_1}\right)\cdot\sum_{n_1=1}^N |u^{(1)}_{n_1}|+o(N),\label{key_step:2}
\end{align}
where step $(a)$ uses the shorthand
\begin{equation}
\eta\triangleq\angle\left(h_{0,0}+\sum^N_{n_2=1}h_{0,n_2}\right)-\angle\left(\sum^N_{n_2=1}u^{(2)}_{n_2}\right).
\end{equation}
The key step in the above derivation is that $\eta$ is independent of $n_1$ and hence can be omitted after step $(a)$. Moreover, step $(a)$ follows by the rank-one assumption in \eqref{eqn 2 decompose} and the definition of $\hat\theta^\star_{n_1}$ in \eqref{eqn 2 theta^}, and step $(b)$ follows by the bound between $\theta'_{n_1}$ and $\hat\theta^\star_{n_1}$  in \eqref{bound:3}. Finally, combining \eqref{key_step:1} and \eqref{key_step:2} gives
\begin{align}
    |g(\theta'_1,\theta'_2)|^2 &= \Omega\Bigg(\Bigg(\sum_{n_2=1}^N\sum_{n_1=1}^N |u^{(2)}_{n_2}||u^{(1)}_{n_1}|\Bigg)^2\Bigg)\notag\\
    &=\delta^2_1\delta^2_2\Omega(N^4).
\end{align}
The proof of Theorem \ref{coro boost 2} is then completed. 
\end{IEEEproof}

\subsection{Comments on Theorem \ref{coro boost 2}}
\label{subsec:prev_work}

Although the conditions C1--C3 are not easy to comprehend at first glance, the intuitions behind them turn out to be quite simple. C1 suggests that the channel between any two adjacent IRSs should not be blocked; this makes sense because the two IRSs reduce to a larger single IRS otherwise. C2 suggests that the number of phase shift choices should be more than two; Example \ref{exp:C2} will show a worst-case scenario with $K_1=K_2=2$ in which the IRS gain is arbitrarily close to zero. C3 suggests that the longest reflected channels should be sufficiently strong; this also makes sense because we wish to make sure that the majority of the channels are good, i.e., there are $N^2$ two-hop channels, while there are only $2N$ one-hop channels and one direct channel.

Furthermore, we compare the proposed conditions with the existing conditions for the double-IRS system to achieve an SNR boost of $\Theta(N^4)$. The conditions in \cite{Han_double_IRS_beamforming_power_scaling} are 
\begin{enumerate}[C'1.]
    \item the matrix factorizing property in C1 holds for all the two-hop reflected channels;
    \item $K_1,K_2\rightarrow\infty$, namely the continuous beamforming;
    \item the direct channel $h_{0,0}$ and the one-hop reflected channels $\{h_{n_1,0},h_{0,n_2}\}$ all equal zero.
\end{enumerate}
It is easy to see that C'2 is a special case of C2 and that C'3 is a special case of C3. Thus, $\{$C'1--C'3$\}\subset\{$C1--C3$\}$.

Another set of conditions for the double-IRS system to reach the $\Theta(N^4)$ SNR boost is proposed in \cite{Han2022} as
\begin{enumerate}[C''1.]
    \item the matrix factorizing property in C1 holds for all the two-hop reflected channels;
    \item $K_1,K_2\rightarrow\infty$, namely the continuous beamforming;
    \item the direct channel $h_{0,0}$ equals zero;
    \item the so-called \emph{array response} \cite{Han2022} of IRS 1 remains the same for the IRS-IRS link and the IRS-receiver link; the array response of IRS 2 remains the same for the IRS-IRS link and the transmitter-IRS link; intuitively, this condition holds if IRS 1 is very close to the transmitter while IRS 2 is very close to the receiver.
\end{enumerate}

It turns out that $\{$C1--C3$\}$ and $\{$C''1--C''3$\}$ are not contained in each other. Nevertheless, the latter only works for the free-space propagation channel model while the former does not assume any particular channel model. Section \ref{sec test} will show that the proposed conditions $\{$C1--C3$\}$ are much easier to satisfy.

As the final part of this section, we illustrate through some concrete examples why the proposed conditions C1--C3 are critical to the $\Theta(N^4)$ boost for a double-IRS system.

\begin{example}[Why is condition C1 needed?]
Assume that $K_1=K_2=4$ and $N$ is an odd number. For any $n_1$ and $n_2$, assume that $h_{0,0}=h_{n_1,0}=h_{0,n_2}=0$, $h_{n_1,n_2}=\beta e^{j(n_1+n_2)\pi}$ if $n_1\neq n_2$, and $h_{n_1,n_2}=\beta e^{j(n_1+n_2)\pi}+2\beta$ if $n_1=n_2$, where $\beta>0$ is a positive constant. This channel setting satisfies C2 and C3 but violates C1 since $h_{n_1,n_2}h_{n_2,n_1}\neq h_{n_1,n_1}h_{n_2,n_2}$ for $n_1\neq n_2$.
It can be shown that the alternating CSM method yields $\theta'_{n_1}=\theta'_{n_2}=0$ for every $n_1$ and $n_2$ in this case. As a result, $|\xi_{n_2}|=O(1)$ and thus $|g(\theta'_{n_1},\theta'_{n_2})|^2$ is $O(N^2)$ according to the first line of \eqref{key_step:1}, so the SNR boost is at most quadratic.

Instead, if we let $h_{n_1,n_2}=\beta e^{j(n_1+n_2)\pi}$ for any $(n_1,n_2)$ in this example, then C1 is satisfied with $u^{(1)}_{n_1}=\sqrt{\beta}e^{jn_1\pi}$ and $u^{(2)}_{n_2}=\sqrt{\beta}e^{jn_2\pi}$. We have $\theta'_{n_1}=0$ if $n_1$ is odd and $\theta'_{n_1}=\pi$ otherwise, and $\theta'_{n_2}=0$ if $n_2$ is odd and $\theta'_{n_2}=\pi$ otherwise. Substituting the above $(\theta'_{n_1},\theta'_{n_2})$ in \eqref{g L=2} gives a $\Theta(N^4)$ boost.
\end{example}

\begin{example}[Why is condition C2 needed?]
\label{exp:C2}
Assume that $K_1=K_2=2$ and $N$ is an odd number. For any $n_1$ and $n_2$, assume that $h_{0,0}=h_{0,n_2}=0$, $u^{(1)}_{n_1}=\sqrt{\beta}e^{j(n_1+\frac{1}{2})\pi}$, and $u^{(2)}_{n_2}=\sqrt{\beta}e^{jn_2\pi}$, where $\beta>0$ is a positive constant. Moreover, let $h_{n_1,0}=\frac{1}{3}\beta e^{j\frac{\pi}{4}}$ for odd $n_1$ and let $h_{n_1,0}=\frac13\sqrt{\beta}e^{-j\frac{\pi}{4}}$ for even $n_1$. Notice that the above setting satisfies all the conditions in Theorem \ref{prop:CSM} except C2. We have $\theta'_{n_1}=0$ for all $n_1$, and $\theta'_{n_2}=0$ if $n_2$ is odd and $\theta'_{n_2}=\pi$ otherwise. The resulting SNR boost is $\Theta(N^2)$.

In contrast, if $K_1$ and $K_2$ are raised to $4$ in this example, then we have $\theta'_{n_1}=-\frac{\pi}{2}$ if $n_1$ is odd and $\theta'_{n_1}=\frac{\pi}{2}$ otherwise, $\theta'_{n_2}=0$ if $n_2$ is odd and $\theta'_{n_2}=\pi$ otherwise. Substituting the above $(\theta'_{n_1},\theta'_{n_2})$ in \eqref{g L=2} gives a $\Theta(N^4)$ boost.
\end{example}

\begin{example}[Why is condition C3 needed?]
\label{exp:C3}
Assume that $K_1=K_2=4$ and $N$ is an odd number. For any $n_1$ and $n_2$, assume that $h_{0,0}=h_{0,n_2}=0$, $h_{n_1,0}=2\beta e^{j\frac{\pi}{2}}$, $u^{(1)}_{n_1}=\sqrt{\beta}e^{jn_1\pi}$ and $u^{(2)}_{n_2}=\sqrt{\beta}e^{jn_2\pi}$,where $\beta>0$ is a positive constant.
Observe that $|h_{n_1,0}|=2|\sum_{n_2=1}^N h_{n_1,n_2}|$ under the above setting, so C3 in Theorem \ref{prop:CSM} does not hold here (but C1 and C2 are satisfied). In this case $\theta'_{n_1}=-\frac{\pi}{2}$ for every $n_1$, and $\theta'_{n_2}=\frac{\pi}{2}$ if $n_2$ is odd and $\theta'_{n_2}=-\frac{\pi}{2}$ otherwise. As a result, the SNR boost is $\Theta(N^2)$ in this case.

In contrast, if $u^{(1)}_{n_1}=u^{(2)}_{n_2}=\sqrt{\beta}$ for every pair $(n_1,n_2)$ in this example, then C3 can be satisfied by letting $\gamma=\frac{\pi}{8}$ when $N$ is sufficiently large. In this case, we have $\theta'_{n_1}=0$ for every $n_1$, and $\theta'_{n_2}=\frac{\pi}{2}$ for every $n_2$. As a result, the SNR boost of $\Theta(N^4)$ is achieved.
\end{example}

\section{General $L$-IRS Case}
\label{sec csm}

\begin{algorithm}[t]
\caption{Blind Beamforming for an $L$-IRS System}
\label{alg:SCSM}
\begin{algorithmic}[1]
\State{Initialize all the $\theta_{n_\ell}$'s to zero.}
\For{$\ell=1,\ldots,L$}
\State{Generate $T$ random samples $\{\theta^{(t)}_{n_\ell}|n_\ell=1,\ldots,N\}$.}
\For{$t=1,\ldots,T$}
\State{Measure the received signal power $|Y^{(t)}|^2$.}
\EndFor
\For{$n_\ell=1,\ldots,N$}
\For{$k=1,\ldots,K_\ell$}
\State{Compute the conditional sample mean in \eqref{cond_expectation}.}
\EndFor
\State Decide each $\theta_{n_{\ell}}$ for IRS $\ell$ according to \eqref{CSM}.
\EndFor
\EndFor
\end{algorithmic}
\end{algorithm} 

The CSM method can be further extended to more than two IRSs in a sequential fashion. The initial values of all the $\theta_{n_\ell}$'s are set to zero. We optimize one IRS at a time while holding the rest IRSs fixed. Algorithm \ref{alg:SCSM} summarizes this sequential CSM method.

Most importantly, the performance bound analysis in Theorem \ref{coro boost 2} can be carried over to the general $L\ge2$ IRSs, as stated in the theorem below.

\begin{theorem}
\label{Proposition:boost}
If an $L$-IRS system satisfies the following three conditions:
\begin{enumerate}[D1.]
    \item there exist a set of nonzero values $\{u^{(\ell)}_{n_l}\in\mathbb C\,|\, n\in[1:N],\;\ell\in[1:L]\}$ such that each $L$-hop channel (which is related to every IRS) can be decomposed as
    \begin{equation}
    \label{decompose}
    h_{n_1,\ldots,n_L}=\prod^L_{\ell=1}u^{(\ell)}_{n_\ell},
    \end{equation}
    where none of $\{n_1,\ldots,n_L\}$ equals zero;
    \item the number of phase shift choices satisfy $K_L\ge3$ and $\sum_{\ell=1}^{L-1}\frac{1}{K\ell}<\frac{1}{2}$;
    \item there exists a constant $\gamma\in[0,\frac{\pi}{L-1}(\frac{1}{2}-\sum_{\ell=1}^{L-1}\frac{1}{K_\ell}))$ such that
    \begin{multline}
        \frac{\sum_{(n_1,\ldots,n_L)\in\mathcal A^{(\ell)}_{m}}\!|h_{n_1,\ldots,n_L}|}{\prod_{i>\ell}\!\big|\!\sum^N_{n_i=1}u^{(i)}_{n_i}\big|\cdot\prod_{i<\ell}\!\big[\!\sum^N_{n_i=1}|u^{(i)}_{n_i}|\cos(\gamma+\frac{\pi}{K_\ell})\big]}\\
        \le |u^{(\ell)}_{m}|\cdot \sin\gamma\label{eqn sin gamma}
    \end{multline}
    for $\ell\in[1:L-1]$ and $m\in[1:N]$, where 
    \begin{equation}
        \mathcal A^{(\ell)}_{m} \triangleq\left\{(n_1,\ldots,n_L)\Bigg|n_\ell=m,\prod^L_{\ell=1}n_\ell=0\right\},
    \end{equation}
\end{enumerate}
then Algorithm \ref{alg:SCSM} yields an SNR boost of $N^{2L}$ as follows:
\begin{align}
\mathbb{E}\left[\frac{\textrm{SNR}}{\textrm{SNR}_0}\right]=\frac{\prod^L_{\ell=1}\delta^2_\ell}{|h_0|^2}\cdot \Theta(N^{2L}),\label{eqn snr boost}
\end{align}
so long as $T=\Omega\left(N^2(\log(NL))^3+N^2L\log(NL)\right)$ and $L=o(N)$, where the expectation is taken over random samples of phase shifts, and
\begin{align}
\delta_\ell\triangleq \frac{1}{N}\sum^N_{n_\ell=1}|u^{(\ell)}_{n_\ell}|.\;\;\forall \ell\in[1:L],\label{eqn delta}
\end{align}
\end{theorem}

The following lemma plays a key role in proving the above theorem.

\begin{lemma}\label{lemma gamma+pi/K}
Let $\theta'_{n_\ell}$ be the decision of $\theta_{n_\ell}$ by Algorithm \ref{alg:SCSM}, and generalize the definition of $\hat\theta^\star_{n_1}$ in \eqref{eqn 2 theta^} as
\begin{multline}
\label{L-IRS:hat_theta_star}
    \hat\theta^\star_{n_\ell}\triangleq\angle\left(\sum_{(m_1,\ldots,m_L)\in\mathcal D^{(\ell)}_{0}}h_{m_1,\ldots,m_L}e^{j\sum^{\ell-1}_{i=1}\theta'_{m_i}}\right)-\angle u^{(\ell)}_{n_\ell}\\
    -\angle\left[\sum_{(m_1,\ldots,m_L)\in\mathcal E^{(\ell)}_{n_\ell}}\left(e^{j\sum^{\ell-1}_{i=1}\theta'_{m_i}}\prod_{i\neq \ell}u^{(i)}_{m_i}\right)\right],
\end{multline}
where
\begin{equation}
    \mathcal D^{(\ell)}_{m} = \left\{(m_1,\ldots,m_L)|m_\ell=m\right\}
\end{equation}
and
\begin{equation}
    \mathcal E^{(\ell)}_{m} = \Bigg\{(m_1,\ldots,m_L)\Bigg|m_\ell=m,\prod_{i\ne\ell}m_{i}\ne0\Bigg\}.
\end{equation}
For any RE $n_\ell$, we have
\begin{equation}
\label{theta_ell:inequality}
    |\hat\theta^\star_{n_\ell}-\theta'_{n_\ell}|\le\gamma+\frac{\pi}{K_\ell}
\end{equation}
given the constant $\gamma$ as defined in the condition D3.
\end{lemma}
\begin{IEEEproof}
See Appendix A.
\end{IEEEproof}

Equipped with the inequality in \eqref{theta_ell:inequality}, we are now ready to prove Theorem \ref{Proposition:boost}. The effective channel from the transmitter to the receiver is written as a function $g:\Phi_K^L\mapsto\mathbb C$ of the beamforming decision $(\theta_{n_1},\ldots,\theta_{n_L})$, i.e.,
\begin{equation}
g(\theta_{n_1},\ldots,\theta_{n_L}) =\!\sum_{(n_1,\ldots,n_L)\in[0:N]^L}\!h_{n_1,\ldots,n_L}e^{j\sum^L_{\ell=1}\theta_{n_{\ell}}}.\label{g}
\end{equation}
To establish an SNR boost of $\Theta(N^{2L})$, it suffices to show that $\mathbb E[|g|^2]=\prod^L_{\ell=1}\delta^2_{\ell}\cdot\Theta(N^{2L})$. Again, the converse $\mathbb E[|g|^2]=\prod^L_{\ell=1}\delta^2_{\ell}\cdot O(N^{2L})$ is evident, so the rest of this section focuses on the achievability $\mathbb E[|g|^2]=\prod^L_{\ell=1}\delta^2_{\ell}\cdot\Omega(N^{2L})$.

Let us first recall how $\theta_{n_\ell}$ is decided in Algorithm \ref{alg:SCSM}. We denote by $\mathcal D^{(\ell)}_{0}$ the set of channels not related to any RE of IRS $\ell$.
When optimizing $\theta_{n_\ell}$, all the channels in $\mathcal D^{(\ell)}_{0}$ are treated as direct channels, while the rest channels are treated as reflected channels. Recall also that all those IRSs $i>\ell$ have not yet been configured when optimizing $\theta_{n_\ell}$, so we have $\theta_{n_i}=0$ for $i\in[\ell+1:L]$. The continuous solution of $\theta_{n_\ell}$ is then given by
\begin{multline}
\theta^\star_{n_\ell} =\angle\Bigg(\sum_{(m_1,\ldots,m_L)\in\mathcal D^{(\ell)}_0}h_{m_1,\ldots,m_L}e^{j\sum^{\ell-1}_{i=1}\theta'_{m_i}}\Bigg)\\
    -\angle\Bigg(\sum_{(m_1,\ldots,m_L)\in\mathcal D^{(\ell)}_{n_\ell}}h_{m_1,\ldots,m_L}e^{j\sum^{\ell-1}_{i=1}\theta'_{m_i}}\!\Bigg).\label{eqn theta star}
\end{multline}
As shown in Appendix B of \cite{blind_beamforming_twc}, the probability that $\theta'_{n_\ell}\neq\arg\min_{\theta\in{\Phi_{K_\ell}}}|\theta-\theta^\star_{n_\ell}|$ for RE $n_\ell$ is bounded above as
\begin{align}
&\mathbb{P}\left\{\theta'_{n_\ell}\neq\arg\min_{\theta\in{\Phi_{K_\ell}}}|\theta-\theta^\star_{n_\ell}|\right\}\notag\\
&\le \Theta\left(\alpha^{\frac{T}{N^2q^2}}\right)+\Theta\left(\frac{qN}{T}\right)+\Theta\left(e^{-q/4}\right),\notag
\end{align}
where $\alpha\in(0,1)$ is a constant related to the channel status and $q$ is a parameter to be specified later. Based on the union of the events bound, the probability that there exists an RE $n_\ell$ in any of the $L$ IRSs such that $\theta'_{n_\ell}\neq\arg\min_{\theta\in{\Phi_{K_\ell}}}|\theta-\theta^\star_{n_\ell}|$ is bounded above as
\begin{align}
&\mathbb{P}\left\{\text{at least one }\theta'_{n_\ell}\neq\arg\min_{\theta\in{\Phi_{K_\ell}}}|\theta-\theta^\star_{n_\ell}|\right\}\notag\\
&\le \sum_{\ell=1}^L \sum_{n_\ell=1}^N \mathbb{P}\left\{\theta'_{n_\ell}\neq\arg\min_{\theta\in{\Phi_{K_\ell}}}|\theta-\theta^\star_{n_\ell}|\right\}\notag\\
&\le \Theta\left(NL\alpha^{\frac{T}{N^2q^2}}\right)+\Theta\left(\frac{qN^2L}{T}\right)+\Theta\left(NLe^{-q/4}\right).\notag
\end{align}
We can further show that the above error probability is bounded by an $o(1)$ term by letting $q=\Omega(\log(NL))$ and $T=\Omega\left(N^2(\log(NL))^3+N^2L\log(NL)\right)$. Thus, so long as $T=\Omega\left(N^2(\log(NL))^3+N^2L\log(NL)\right)$, Algorithm \ref{alg:SCSM} would round the above continuous solution to the discrete set $\Phi_{K_\ell}$:
\begin{equation}
\theta'_{n_\ell}=\arg\min_{\theta\in{\Phi_{K_\ell}}}|\theta-\theta^\star_{n_\ell}|
\end{equation}
and hence
\begin{equation}
|\theta'_{n_\ell}-\theta^\star_{n_\ell}|\le \frac{\pi}{K_\ell}.\label{eqn theta' star}
\end{equation}

We then bound the channel strength as follows:
\begin{align}
    &|g(\theta'_{n_1},\ldots,\theta'_{n_L})|^2\notag\\
    &=\Bigg|\sum_{(n_1,\ldots,n_L)\in[0:N]^L}h_{n_1,\ldots,n_L}e^{j\sum_{i=1}^{L}\theta'_{n_i}}\Bigg|^2\notag\\
    &= \Bigg|\sum^N_{n_L=0}e^{j\theta'_{n_L}}\Bigg(\sum_{(n_1,\ldots,n_{L-1})}h_{n_1,\ldots,n_L}e^{j\sum_{i=1}^{L-1}\theta'_{n_i}}\Bigg)\Bigg|^2\notag\\
    &\overset{(a)}{\ge} \left[\sum^N_{n_L=1}\!\cos(\theta'_{n_L}-\theta^\star_{n_L}\!)\left|\sum_{(n_1,\ldots,n_{L-1})}\!\! h_{n_1,\ldots,n_L}e^{j\sum_{i=1}^{L-1}\theta'_{n_i}}\right|\right]^2\notag\\
    &\overset{(b)}{\ge} \left[\sum^N_{n_L=1}\cos\left(\frac{\pi}{K_L}\right)\left|\sum_{(n_1,\ldots,n_{L-1})}h_{n_1,\ldots,n_L}e^{j\sum_{i=1}^{L-1}\theta'_{n_i}}\right|\right]^2\label{eqn g2 1 1},
\end{align}
where step $(a)$ follows by only considering the projection of every $e^{j\theta'_{n_L}}\sum_{(n_1,\ldots,n_{L-1})}h_{n_1,\ldots,n_L}e^{j\sum_{i=1}^{L-1}\theta'_{n_i}}$ with $n_L\ne0$ onto that with $n_L=0$, and step $(b)$ follows by the inequality in \eqref{eqn theta' star} directly. 

For a fixed $n_L\ne0$, we can further bound the sum component in \eqref{eqn g2 1 1} as follows:
\begin{align}
&\left|\sum_{(n_1,\ldots,n_{L-1})\in[0:N]^{L-1}}h_{n_1,\ldots,n_L}e^{j\sum_{i=1}^{L-1}\theta'_{n_i}}\right|-o(N^{L-1})\notag\\
&\overset{(a)}{=}\left|\sum_{(n_1,\ldots,n_{L})\in\mathcal E^{(L)}_{n_L}}h_{n_1,\ldots,n_L}e^{j\sum_{i=1}^{L-1}\theta'_{n_i}}\right|\notag\\
&=\left|\sum_{(n_1,\ldots,n_{L})\in\mathcal E^{(L)}_{n_L}}\left(u^{(L)}_{n_L}\prod^{L-1}_{\ell=1}u^{(\ell)}_{n_\ell}e^{j\theta'_{n_\ell}}\right)\right|\notag\\
&\overset{(b)}{=}\big|u^{(L)}_{n_L}\big|\cdot\left|\sum_{(n_1,\ldots,n_{L})\in\mathcal E^{(L)}_{n_L}}\prod^{L-1}_{\ell=1}\big|u^{(\ell)}_{n_\ell}\big|e^{j(\theta'_{n_\ell}-\hat\theta^\star_{n_\ell})}\right|\notag\\
&\overset{(c)}{\ge}\big|u^{(L)}_{n_L}\big|\sum_{(n_1,\ldots,n_{L})\in\mathcal E^{(L)}_{n_L}}\cos\left[(L-1)\gamma+\sum_{\ell=1}^{L-1}\frac{\pi}{K_\ell}\right]\prod^{L-1}_{\ell=1}\big|u^{(\ell)}_{n_\ell}\big|\notag\\
&=|u^{(L)}_{n_L}|N^{L-1}\cos\left[(L-1)\gamma+\sum_{\ell=1}^{L-1}\frac{\pi}{K_\ell}\right]\prod^{L-1}_{i=1}\delta_i\label{eqn each term 6},
\end{align}
where step $(a)$ follows as the number of $(n_1,\ldots,n_{L-1})$'s with at least one $n_i=0$ equals $(N+1)^{L-1}-N^{L-1}=o(N^{L-1})$, step $(b)$ follows by \eqref{L-IRS:hat_theta_star}, and step $(c)$ follows by \eqref{theta_ell:inequality}.

\begin{figure}[t]
\centering
\includegraphics[width=8.0cm]{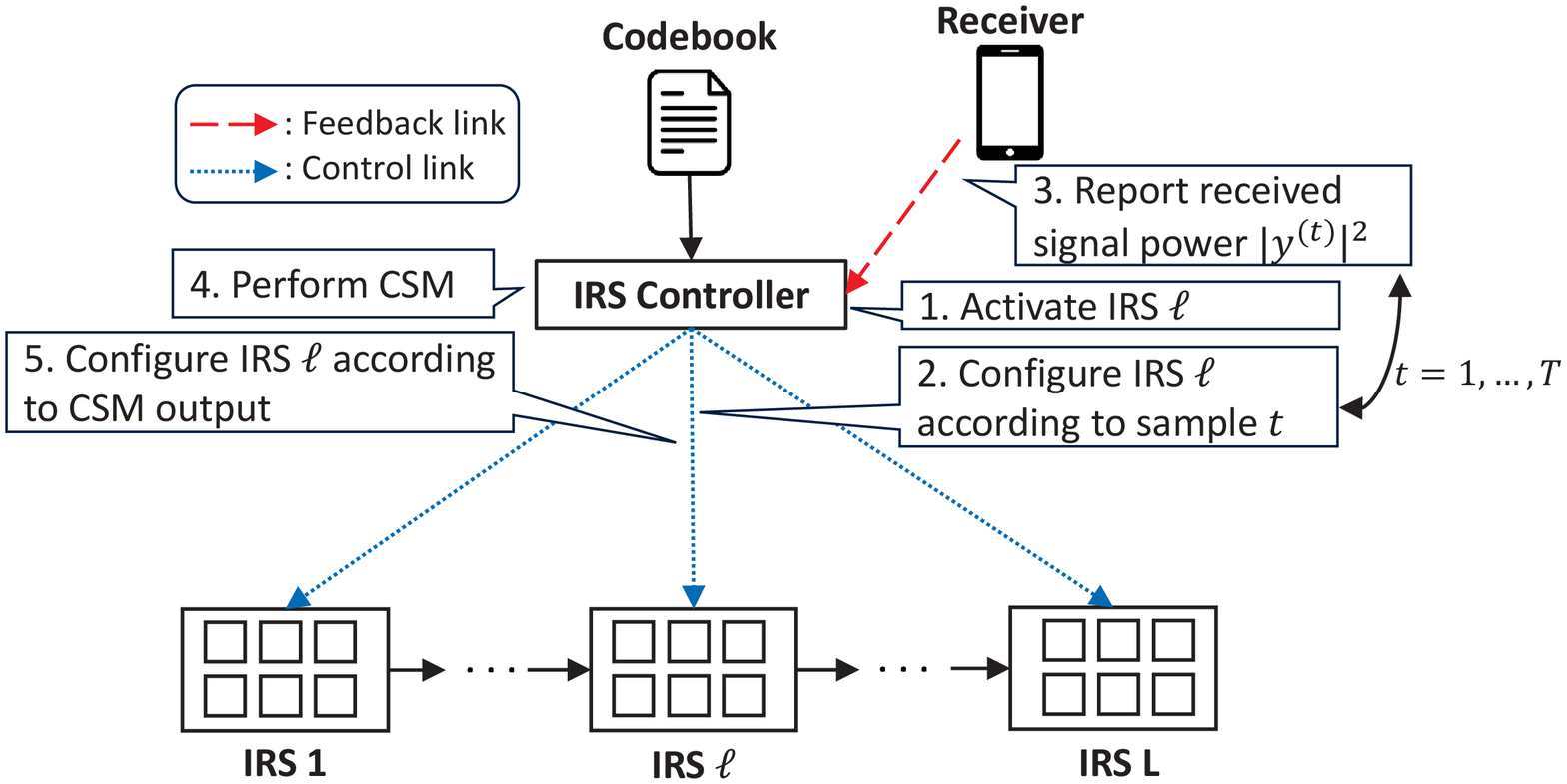}\\
\vspace{1em}
\small{(a) controller-centered type}\\
\vspace{2em}
\includegraphics[width=8.5cm]{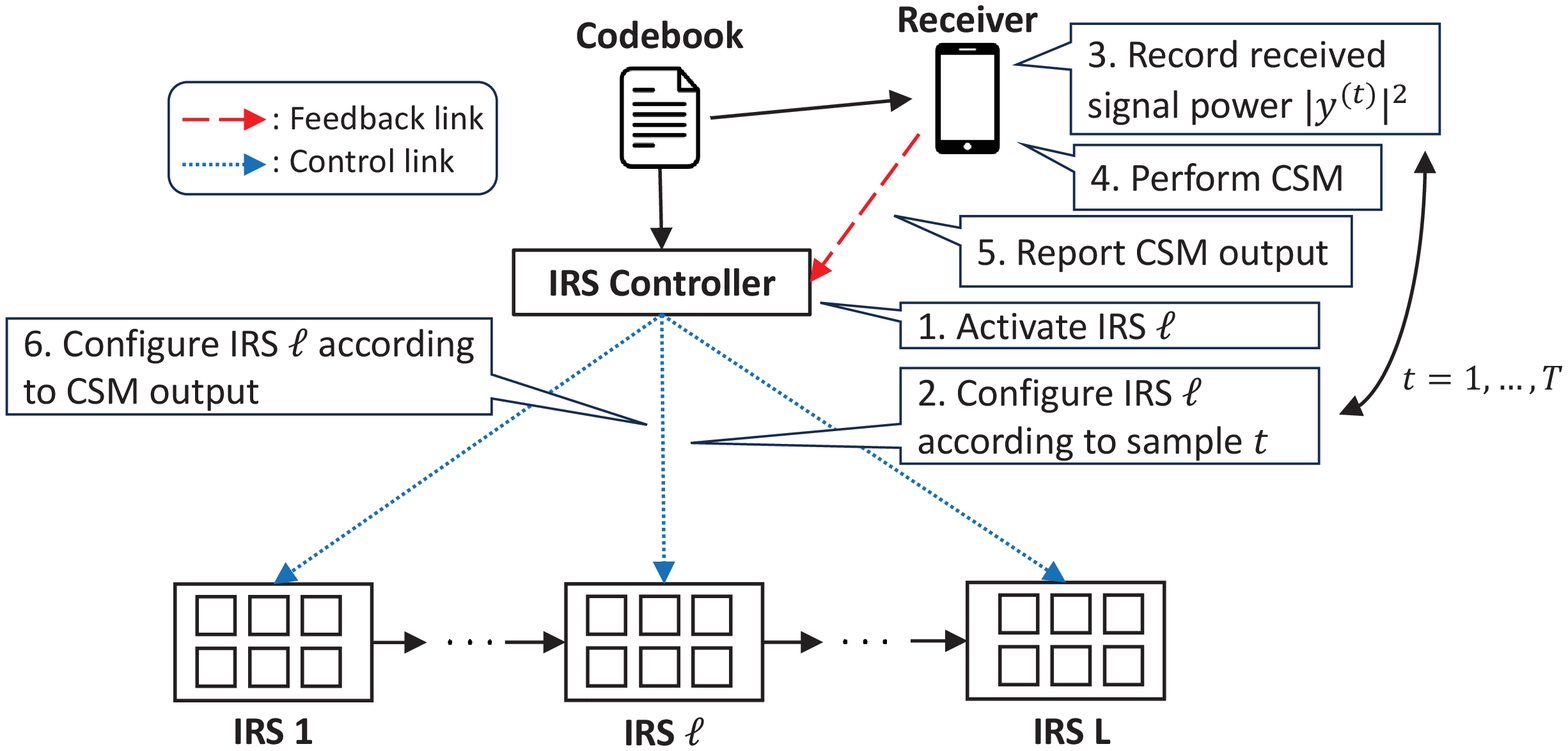}\\
\vspace{1em}
\small{(b) receiver-centered type}
\caption{Two different paradigms of blind beamforming.}
\label{fig:paradigm}
\end{figure}

Substituting the lower bound \eqref{eqn each term 6} in \eqref{eqn g2 1 1}, we obtain
\begin{equation}
|g(\theta'_{n_1},\ldots,\theta'_{n_L})|^2=\left(\prod^L_{\ell=1}\delta^2_\ell\right)\cdot\Omega(N^{2L}).
\end{equation}
Furthermore, combining the above result with the evident fact that $\mathbb E[|g|^2]=\prod^L_{\ell=1}\delta^2_{\ell}\cdot O(N^{2L})$ leads us to the boost $\Theta(N^{2L})$. The proof of Theorem \ref{Proposition:boost} is thus completed. $\hfill\blacksquare$

We cannot provide a performance guarantee for the blind beamforming method if the proposed conditions are not satisfied. However, we clarify that the proposed conditions are already much easier to satisfy than the existing ones. Besides, our field tests in the next section show that the proposed blind beamforming method still yields excellent performance in the real world even if the conditions are not satisfied. Most importantly, the insights and intuitions behind our theoretical conditions can guide the IRS deployment in practice, so it is worthwhile to expand the existing conditions even if the new conditions still cannot be strictly satisfied.

Theorem \ref{Proposition:boost} implies that only one round of configuration (i.e., every IRS is optimized one time regardless of $L$) suffices to attain an SNR boost of $\Theta(N^{2L})$. This is of practical significance when the IRSs are extensively deployed in the network. We also clarify that the conditions D1--D3 are \emph{sufficient but not necessary} for the $\Theta(N^{2L})$ SNR boost.

We further discuss how the sequential CSM algorithm can be carried out in practice. There are four parties---transmitter, receiver, IRSs, and an IRS controller in the protocol; there is a feedback link from the receiver to the IRS controller; there is a control link from IRS controller to each IRS. A codebook of prescribed pseudo-random samples of phase shift arrays is generated beforehand. There are two possible paradigms of blind beamforming as illustrated in Fig. \ref{fig:paradigm}, which are the controller-centered type and the receiver-centered type.

For the controller-centered paradigm, the codebook is only revealed to the IRS controller. Let us consider the $\ell$th iteration of blind beamforming. First, the IRS controller activates IRS $\ell$ through the control link, and then sets its phase shift array to the samples in the codebook sequentially. In the meanwhile, for each sample, the receiver reports the received signal power to the IRS controller through the feedback link. After completing $T$ samples, the IRS controller uses the power feedback from the receiver so far to perform CSM. Finally, the IRS controller configures IRS $\ell$ as the CSM output through the control link.

Alternatively, by the receiver-centered paradigm, the codebook is revealed to the receiver in addition. The sampling part follows the controller-centered case, but the receiver now records its measured received signal power with respect to each sample, instead of reporting it back to the IRS controller. After $T$ samples, the receiver performs CSM based on the codebook and the received signal power data, and then reports the CSM output to the IRS controller through the feedback link. Finally, the IRS controller configures IRS $\ell$ according to the feedback from the receiver.

\begin{remark}[Same Phase-Shift Resolution]
If the $L$ IRSs have the same phase-shift resolution, i.e., when all the $K_\ell$ values are equal, then D2 in Theorem \ref{Proposition:boost} reduces to every $K_\ell>2(L-1)$.
\end{remark}

\begin{remark}[Overhead Cost]
For the double-IRS case, our theoretical lower bound on $T$ with provable performance is $T=\Omega\left(N^2(\log N)^3\right)$ as stated in Theorem 1; but the practical choice of $T$ can be much smaller as shown in Fig.~\ref{fig:boostvsn}. By contrast, since there are $N^2$ unknown channels, it requires at least $N^2$ pilot symbols to solve the channel estimation equation \cite{channel_est_MIMO_MU_zbx,channel_est_beamforming_ycs}. (The work in \cite{Benicio2021} argues that $2N$ pilot symbols are sufficient for channel estimation for some special cases.) We will explicitly compare blind beamforming with the channel-estimation-based methods \cite{channel_est_beamforming_ycs,Benicio2021} in Section \ref{section simulation}.
\end{remark}

\begin{remark}[Computational Complexity]
For each IRS, it requires $O(NT)$ to compute all the conditional sample mean values, and it further requires $O(NK)$ to decide phase shifts for the current IRS based on the conditional sample mean. For a total of $L$ IRSs, the overall computational complexity of blind beamforming equals $O(LN(K+T))$.
\end{remark}

\section{Experiments}
\label{sec test}

\begin{figure}[t]
\centering
\includegraphics[width=0.45\textwidth]{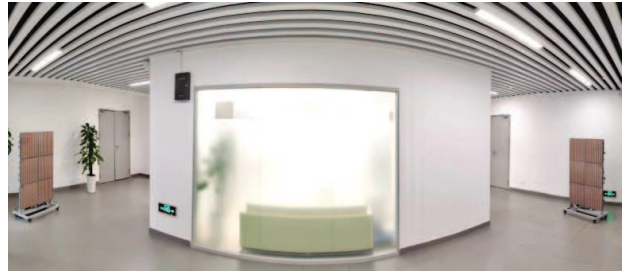}
\caption{Field test with two IRSs deployed in an indoor hallway.}
\label{fig:indoor}
\end{figure}

\begin{figure}[t]
	\centering
    \vspace{-2em}
	\includegraphics[width=7.0cm]{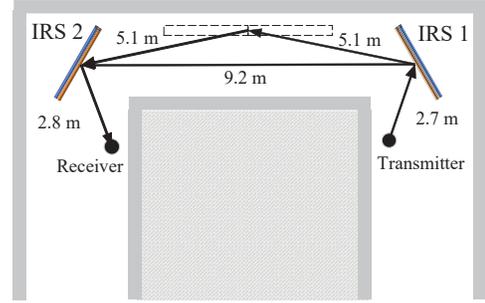}
	\vspace{-1em}
	\caption{Layout drawing of the indoor field test. The two IRSs are placed in two corners for most methods, but are merged into a single larger IRS placed in the middle for ``Physical Single-IRS'' as indicated by the dashed lines.}
	\label{fig:indoor_layout}
\end{figure}

\subsection{Field Tests}
\label{subsec:field_tests}
Throughout our field tests, the transmit power is fixed at $-5$ dBm and the carrier frequency is $2.6$ GHz. The following three IRSs are used:
\begin{itemize}
    \item IRS 1 with 294 REs and 2 phase shift choices $\{0,\pi\}$ for each RE, i.e., $N=294$ and $K_1=2$;
    \item IRS 2 also with $N=294$ and $K_2=2$;
    \item IRS 3 with $N=64$ and $K_3=4$.
\end{itemize}

\begin{figure}[t]
    \centering
    \includegraphics[width=9cm]{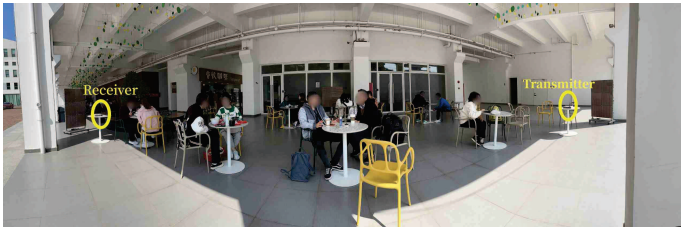}
    \caption{Outdoor field test with three IRSs deployed alongside an open caf\'{e}.}
    \label{fig:outdoor_environment}
\end{figure}

\begin{figure}[t]
    \centering
    \includegraphics[width=6.0cm]{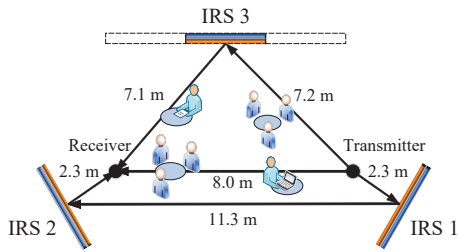}
    \caption{Layout drawing of the outdoor field test. For Physical Single-IRS, we move IRS 1 and IRS 2 to the positions indicated by the dashed lines.}
    \label{fig:outdoor_locations}
\end{figure}

Notice that the condition C2 or D2 is not satisfied in our field test. Actually, the existing IRS prototypes mostly adopt $K=2$ or $K=4$ because of the hardware cost and the system robustness \cite{Dai2020,Trichopoulos2022,pei2021ris,Staat_2022_IRShield}. Notice also that we do not always assume that all the IRSs have the same values of $N$ as in the theoretical model in Section \ref{sec:model}. The following methods are compared:
\begin{itemize}
    \item \emph{Without IRS:} IRS is not used.
    \item \emph{Zero Phase Shifts:} Fix all phase shifts to be zero.
    \item \emph{Random Beamforming:} 
    Try out $L\times1000$ random samples of phase shift vectors and choose the best.
    \item \emph{Virtual Single-IRS:} Ignore the multi-hop channels and treat multiple IRSs as a single one; optimize phase shifts by the method in \cite{blind_beamforming_twc} with $L\times1000$ random samples.
    \item \emph{Physical Single-IRS:} Put multiple IRSs together at the same position to form a single larger IRS; optimize phase shifts by the method in \cite{blind_beamforming_twc} with $L\times1000$ random samples. 
    \item \emph{Proposed Blind Beamforming:} Coordinate multiple IRSs by Algorithm \ref{alg:SCSM} that uses $1000$ random samples per IRS.
\end{itemize}
The SNR boost is evaluated by taking ``without IRS'' as baseline. We consider the following two transmission scenarios:
\begin{itemize}
    
    \item \emph{Indoor Environment:} Deploy IRS 1 and IRS 2 in a U-shaped hallway inside an office building as shown in Fig.~\ref{fig:indoor}. The testbed layout is specified in Fig.~\ref{fig:indoor_layout}. The transmission is blocked by the walls.
    \item \emph{Outdoor Environment:} Deploy three IRSs alongside an open caf\'{e} as shown in Fig.~\ref{fig:outdoor_environment}. The testbed layout is specified in Fig.~\ref{fig:outdoor_locations}. The transmission is occasionally blocked by the crowd and also suffers interference which is treated as noise.
    
\end{itemize}

TABLE \ref{tab:SNR_boost} summarizes the SNR boost performance of the different methods. As shown in the row of Zero Phase Shifts, placing IRSs in the environment (either indoor or outdoor) can already increase SNR by more than 2 dB even without any optimization. Then a simple heuristic optimization method such as Random Beamforming can reap a higher SNR gain. Observe also that Virtual Single-IRS outperforms the above methods significantly, e.g., it improves upon Random Beamforming by around 7 dB for the indoor case. 

In contrast, the proposed Blind Beamforming enhances SNR
further even though C2 or D2 is not satisfied, e.g., its SNR boost is about 5 dB higher than that of
Virtual Single-IRS for the indoor case, and about 3 dB higher
for the outdoor case. This further gain is due to the capability
of Blind Beamforming to take those multi-hop reflections into
account. For this reason, the advantage of Blind Beamforming over Virtual Single-IRS is greater for the indoor case in which the multi-hop reflections play a key role. Another interesting fact from TABLE \ref{tab:SNR_boost} is that Physical Single-IRS is much worse than the other methods especially in the indoor environment. Although its phase shifts have been carefully optimized by the method in \cite{blind_beamforming_twc}, its performance is still limited by the deficiency of multi-hop reflections.

\begin{figure}[t]
    \centering
    \includegraphics[width=6cm]{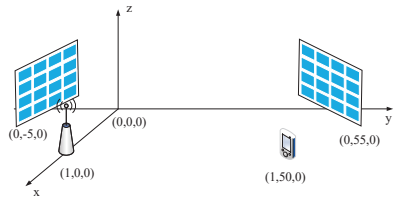}
    \caption{The double-IRS system considered in our simulations. The position coordinates are in meters.}
    \label{fig:double-IRS}
\end{figure}

\begin{table}[t]
\small
    \renewcommand{\arraystretch}{1.3}
\centering
\caption{\small SNR Boosts Achieved by the Different Methods}
\begin{tabular}{lrrr}
\firsthline
& \multicolumn{2}{c}{SNR Boost (dB)} \\
\cline{2-3}
Method       & Indoor & Outdoor   \\
\hline
Zero Phase Shifts    & 2.74  & 2.91 \\
Random Beamforming   & 5.33  & 8.48 \\
Virtual Single-IRS  & 12.07  & 10.80 \\
Physical Single-IRS   & 3.31  & 7.06 \\
Blind Beamforming    & 17.08 & 14.09 \\
\lasthline
\end{tabular}
\label{tab:SNR_boost}
\end{table}

\subsection{Simulation Tests}\label{section simulation}

We now validate the performance of the proposed blind beamforming algorithm in simulations which can admit many more REs at each IRS and many more IRSs. The channels are generated as follows. We refer to the transmitter as node $0$, a total of $L$ IRSs as nodes $1$ through $L$, and the receiver as node $L+1$. We denote by $d_{ij}$ (in meters) the distance between node $i$ and node $j$. The pathloss model follows previous works \cite{Wu2019,Jiang2021a,blind_beamforming_twc}. If there is LoS propagation between node $i$ and node $j$ then the corresponding pathloss is computed as
\begin{align}
    \text{PL}_{ij}=10^{-(30+22\log_{10} (d_{ij}))/20}.
\end{align}
Otherwise, i.e., when the channel between node $i$ and node $j$ is non-line-of-sight (NLoS), the pathloss is computed as
\begin{align}
    \text{PL}_{ij}=10^{-(32.6+36.7\log_{10} (d_{ij}))/20}.
\end{align}
Following \cite{multi_IRS_Mei,Multi_IRS_MIMO_Mei}, we assume that REs are arranged as a uniform linear array with spacing $\xi=0.03$ meters at each IRS. We let the signal wavelength be $\lambda=0.06$ meters. Moreover, denote by $\vartheta_{i,j}$ the angle of departure (AoD) from node $i$ to node $j$, and $\psi_{i,j}$ the angle of arrival (AoA) from node $j$ to node $i$. For the LoS case, the channel from the transmitter to RE $n_\ell$ is given by
\begin{align}
    g_{0,n_\ell}=\sqrt{ \text{PL}_{0,\ell}}\cdot e^{-j\frac{2\pi}{\lambda}d_{0,\ell}} \cdot e^{-j\frac{2\pi}{\lambda}\xi(n_\ell-1)\cos \psi_{\ell,0}},
\end{align}
the channel from RE $n_\ell$ to the receiver is given by
\begin{align}
    &g_{n_\ell,L+1}=\sqrt{ \text{PL}_{\ell,L+1}}\cdot e^{-j\frac{2\pi}{\lambda}d_{\ell,L+1}}\cdot e^{-j\frac{2\pi}{\lambda}\xi(n_\ell-1)\cos \vartheta_{\ell,L+1}},
\end{align}
and the channel from RE $n_\ell$ to RE $n_{\ell'}$, $\ell\ne\ell'$, is given by
\begin{multline}
    g_{n_\ell,n_{\ell'}}=\sqrt{ \text{PL}_{\ell,{\ell'}}}\cdot e^{-j\frac{2\pi}{\lambda}d_{\ell,{\ell'}}} \cdot e^{-j\frac{2\pi}{\lambda}\xi(n_\ell-1)\cos \vartheta_{\ell,{\ell'}}}\\  
    \cdot e^{-j\frac{2\pi}{\lambda}\xi(n_{\ell'}-1)\cos \psi_{\ell',\ell}}.
\end{multline}
For the NLoS case, we generate channels as
\begin{align}
    g_{0,n_\ell} &= \sqrt{ \text{PL}_{0,\ell}}\cdot \zeta_{0,n_\ell},\\
    g_{n_\ell,L+1} &= \sqrt{ \text{PL}_{\ell,L+1}}\cdot\zeta_{\ell,L+1},\\
    g_{n_\ell,n_{\ell'}} &= \sqrt{ \text{PL}_{\ell,{\ell'}}}\cdot\zeta_{n_\ell,n_{\ell'}},
\end{align}
where $\zeta_{0,n_\ell},\zeta_{\ell,L+1},\zeta_{n_\ell,n_{\ell'}}$ are drawn i.i.d. from the complex Gaussian distribution $\mathcal{CN}(0,1)$. For both LoS and NLoS cases, each multi-hop channel $h_{n_1,\ldots,n_L}$ can be obtained by multiplying together a subset of channels in $\{g_{0,n_\ell},g_{n_\ell,L+1},g_{n_\ell,n_{\ell'}}\}$.
Moreover, the transmit power equals $30$ dBm, the background noise power equals $-98$ dBm, and all the $K_\ell$'s are fixed to be $4$ throughout the simulation tests, i.e., $\Phi_{K_\ell} = \left\{ 0 , \frac{\pi}{2}, \pi, \frac{3 \pi}{2}\right\}$ for all $\ell\in[1:L]$.

\begin{figure}[t]
\centering
\includegraphics[width=9cm]{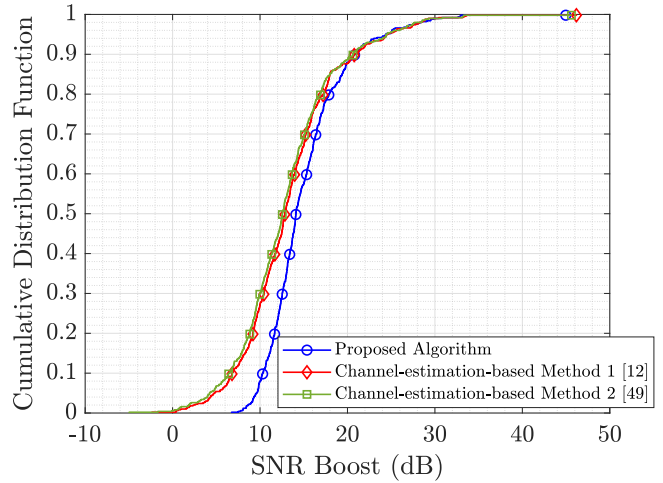}\\
\small{(a) when only the two-hop reflected channels are LoS}\\
\vspace{1em}
\includegraphics[width=9cm]{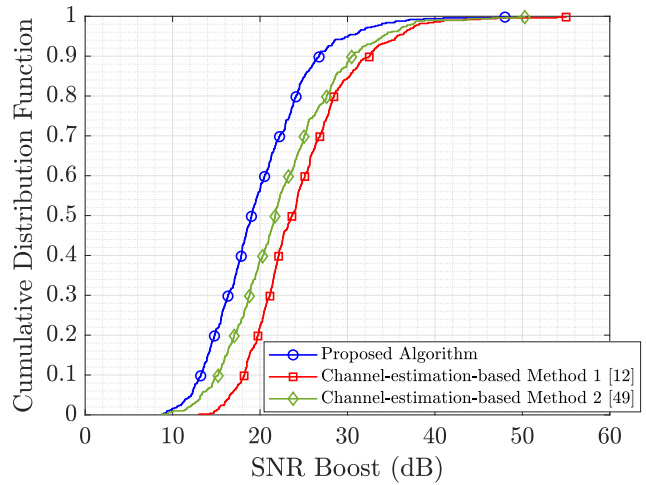}\\
\vspace{1em}
\small{(b) when only the direct channel is NLoS}\\
\caption{Blind beamforming versus channel-estimation-based methods for a double-IRS system.}\label{fig:comparison_CE}
\vspace{-1em}
\end{figure}

\begin{figure}[t]
\centering
\includegraphics[width=9cm]{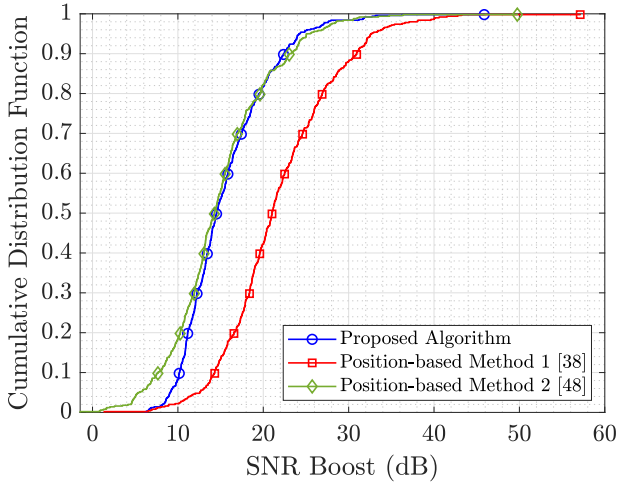}\\
\small{(a) when only the two-hop reflected channels are LoS}\\
\vspace{1em}
\includegraphics[width=9cm]{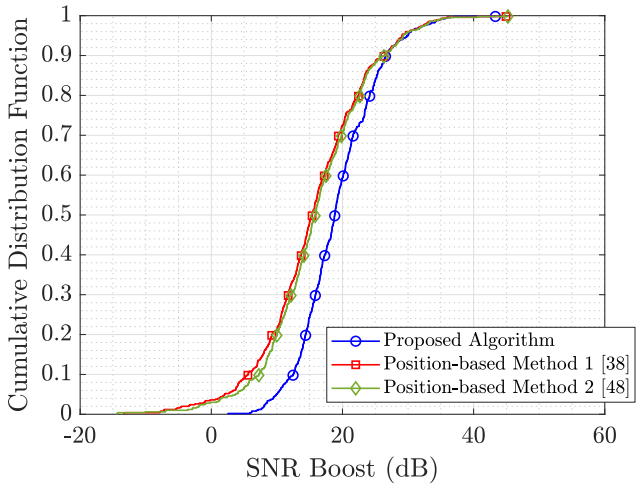}\\
\vspace{1em}
\small{(b) when only the direct channel is NLoS}\\
\caption{Blind beamforming versus position-based methods for a double-IRS system.}\label{fig:comparison_Han}
\end{figure}

\subsubsection{Double-IRS System}
We begin with a double-IRS system as shown in Fig.~\ref{fig:double-IRS}. Assume that the two IRSs have $100$ REs each and that $T=1000$ random samples are taken for each IRS. Assume that $\vartheta_{n_1,n_2}=\vartheta_{n_2,L+1}=\psi_{0,n_1}=\psi_{n_1,n_2}= 5.6^\circ$. Fig. \ref{fig:comparison_CE} compares the performance of our blind beamforming scheme with the channel-estimation-based methods in \cite{channel_est_beamforming_ycs,Benicio2021}. Two different channel scenarios are considered here: Fig. \ref{fig:comparison_CE}(a) assumes that only the two-hop reflected channels are LoS while the rest channels are NLoS; Fig. \ref{fig:comparison_CE}(b) assumes that most channels are LoS except the direct channel. It can be seen that the proposed blind beamforming method outperforms the channel-estimation-based methods in the first scenario; however, when more channels become LoS as in the second scenario, the channel-estimation-based methods perform better. The reason is that if most channels are NLoS then they are too weak to be estimated accurately.

Also, these state-of-the-art channel estimation methods require setting each phase shift according to the DFT matrix, but this can violate the discrete phase shift constraint, e.g., they do not work for the case where each phase shift is either $0$ or $\pi$.
Besides, it is difficult to extend these channel estimation methods to account for more than two IRSs.

Moreover, if $T$ falls below the above bound, our blind beamforming method still works only that its performance cannot be guaranteed by our current theory anymore. However, if there are fewer than $N^2$ pilot symbols, then the linear system for channel estimation cannot be solved, so the channel estimation cannot be performed at all (unless under some certain sparsity assumptions). In our field tests, letting $T=1000L$ can already yield good performance for blind beamforming with $N=294$, so the corresponding overhead is $1000\times2\times2=4000$ symbols, while the estimation methods in \cite{channel_est_MIMO_MU_zbx,channel_est_beamforming_ycs} require at least $294\times294\approx860000$ symbols.

Last, from a practical implementation point of view, the channel estimation methods require reading the received symbol $Y\in\mathbb C$ from the communication chip of the receiver device. By contrast, blind beamforming only requires the received signal power information, which needs a lower level of authority and hence is much easier to implement.

Fig.~\ref{fig:comparison_Han} compares the blind beamforming method with the position-based methods \cite{Han_double_IRS_beamforming_power_scaling,Han2022}. Again two scenarios are considered: Fig.~\ref{fig:comparison_Han}(a) assumes that only the two-hop reflected channels are LoS while Fig.~\ref{fig:comparison_Han}(b) assumes that only the direct channel is NLoS. In the first scenario, we see that blind beamforming is close to the position-based method in \cite{Han_double_IRS_beamforming_power_scaling}, but it is much worse than the method in \cite{Han2022}. Nevertheless, when it comes to the second scenario, blind beamforming outperforms the benchmark methods significantly. It is worth pointing out that these position-based methods \cite{Han_double_IRS_beamforming_power_scaling,Han2022} heavily rely on the \emph{free-space line-of-sight propagation} channel model, while our condition does not assume any particular channel model.

\begin{figure}[t]
    \centering
    \includegraphics[width=9cm]{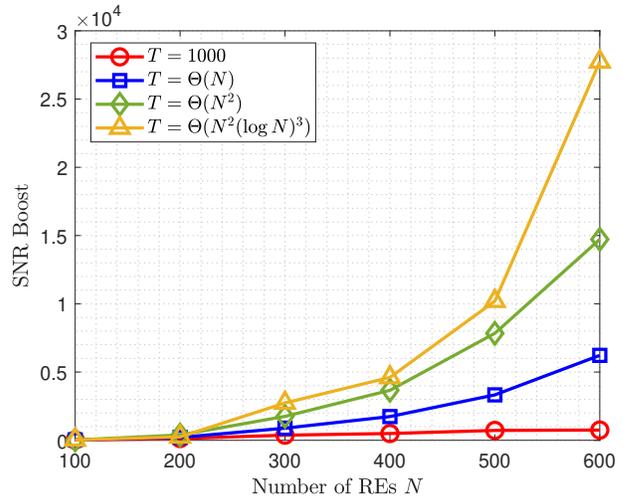}
    \caption{SNR boost versus number of REs under the different $T$ settings.}
    \label{fig:boostvsn}
    \vspace{-1em}
\end{figure}

Moreover, Fig.~\ref{fig:boostvsn} compares the growth curves of the SNR boost with the IRS size $N$ under the different values of $T$. Observe that the growth curve is approximately quartic when $T=\Theta(N^2(\log N)^3)$; this result agrees with Theorem \ref{coro boost 2}. Observe also that the SNR boost still grows rapidly with $N$ even if $T$ falls below the lower bound in Theorem 1, e.g., when $T=\Theta(N^2)$.

\begin{figure}[t]
    \centering
    \includegraphics[width=9cm]{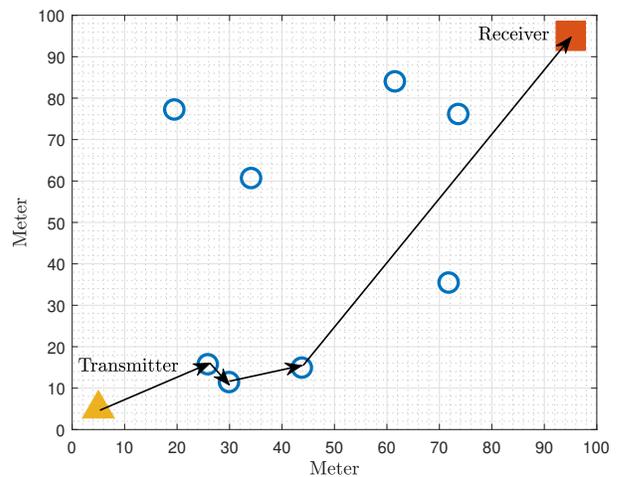}
    \caption{The IRS routing when $L=8$ and $N=600$.}
    \label{fig:routing_n=600}
\end{figure}

\subsubsection{General $L$-IRS System}
We now add more IRSs to the system. Consider a $100\times100$ m$^2$ 2D square area as shown in Fig.~\ref{fig:routing_n=600}. The transmitter is located at $(5,5)$ and the receiver is located at $(95,95)$. There are $8$ possible positions to deploy IRSs: $(19.5,77.3)$, $(25.8,15.8)$, $(29.9,11.5)$, $(34.1,60.7)$, $(43.8,15.0)$, $(61.5,84.1)$, $(71.8,35.5)$ and $(73.6,76.2)$, all in meters. Thus, including the transmitter and receiver, there are a total of $10$ nodes in our case. Following \cite{multi_IRS_Mei}, the propagation status between any nodes are randomly set to LoS with $60\%$ probability and to NLoS with $40\%$ probability. The realization of the propagation statuses in our case can be expressed in an adjacency matrix as
\begin{align*}
    A=
    \begin{bmatrix}
        0 & 0 & 1 & 0 & 0 & 0 & 1 & 0 & 0 & 0 \\
        0 & 0 & 1 & 0 & 1 & 1 & 0 & 0 & 0 & 1 \\
        1 & 1 & 0 & 1 & 0 & 1 & 0 & 1 & 0 & 0 \\
        0 & 0 & 1 & 0 & 1 & 1 & 1 & 0 & 0 & 1 \\
        0 & 1 & 0 & 1 & 0 & 1 & 1 & 1 & 0 & 1 \\
        0 & 1 & 1 & 1 & 1 & 0 & 1 & 1 & 0 & 1 \\
        1 & 0 & 0 & 1 & 1 & 1 & 0 & 1 & 1 & 0 \\
        0 & 0 & 1 & 0 & 1 & 1 & 1 & 0 & 1 & 0 \\
        0 & 0 & 0 & 0 & 0 & 0 & 1 & 1 & 0 & 1 \\
        0 & 1 & 0 & 1 & 1 & 1 & 0 & 0 & 1 & 0 
    \end{bmatrix},
\end{align*}
where the entry $A_{ij}$ equals $1$ if the channel between node $i$ and node $j$ is LoS, and $0$ otherwise.

\begin{figure}[t]
    \centering
    \includegraphics[width=9cm]{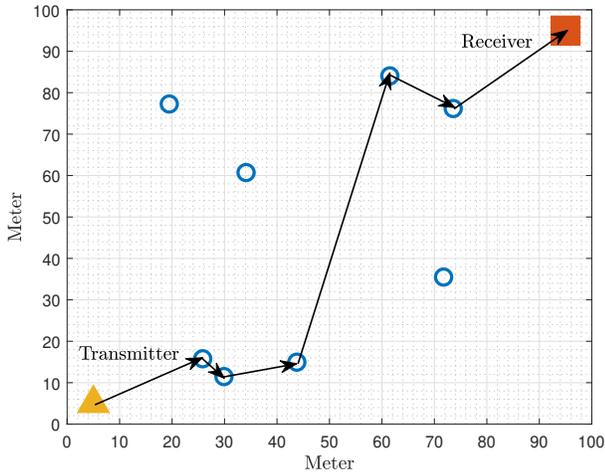}
    \caption{The IRS routing when $L=8$ and $N=1200$.}
    \label{fig:routing_n=1200}
    \vspace{-1em}
\end{figure}

\begin{figure}[t]
    \centering
    \includegraphics[width=9cm]{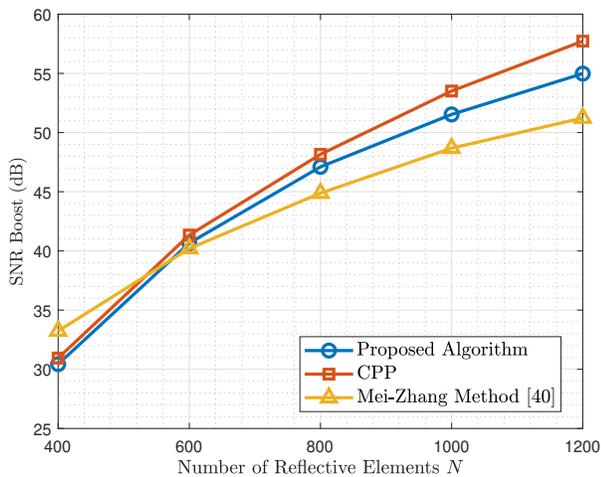}
    \caption{SNR boost versus number of REs for the different algorithms.}
    \label{fig:comparison_beamforming_routing}
\end{figure}

As suggested in \cite{multi_IRS_Mei}, only a subset of the possible positions are selected for the IRS placement, namely the IRS routing. For instance, Fig.~\ref{fig:routing_n=600} shows the IRS placement when $N=600$, and Fig.~\ref{fig:routing_n=1200} shows the IRS placement when $N=1200$, both based on the graph-theoretical algorithm in \cite{multi_IRS_Mei}. Observe that more IRSs are used for routing when the number of REs per IRS $N$ is increased. The rationale is that we have a tradeoff when deciding how many IRSs to use. Clearly, using more IRSs gives more multi-hop reflected paths; in order to reap the multi-path gain, we need to orient the reflection beam of each IRS toward the subsequent IRS, but this requires high-resolution beamforming and thus $N$ should be sufficiently large. On the other hand, using too many IRSs would increase the reflection loss. Due to the above two factors, the number of IRSs should be neither too large nor too small. Now, if $N$ becomes larger, then the first factor outweighs, so it encourages using more IRSs. Actually, this tradeoff has already been pointed out in the previous work \cite{multi_IRS_Mei}.

For the comparison purpose, we consider the following two beamforming methods as benchmarks:
\begin{itemize}
    \item \textit{CPP with Perfect CSI}: Assume that the precise channel information is already known. Then perform CPP across the IRSs sequentially.
    \item \textit{Mei-Zhang Method with Perfect CSI} \cite{multi_IRS_Mei}: First perform the continuous beamforming algorithm in \cite{multi_IRS_Mei} and then round the solution to the discrete set $\left\{ 0, \frac{\pi}{2}, \pi, \frac{3\pi}{2} \right\}$. This method assumes that perfect CSI is available.
\end{itemize}
Notice that the above two competitor methods both require perfect CSI. However, to the best of our knowledge, there is yet no effective way of estimating channels for more than 2 IRSs in the existing literature.

Fig.~\ref{fig:comparison_beamforming_routing} compares the proposed Algorithm \ref{alg:SCSM} with the benchmarks under the different settings of $N$. For Algorithm \ref{alg:SCSM}, we let $T=20\times N$. It can be seen that the proposed algorithm is fairly close to CPP. Actually, the proposed algorithm would approach CPP when $T\rightarrow\infty$; the former can be thought of as a practical implementation of the latter. Mei-Zhang method is about 2 dB higher than the proposed algorithm when $N=400$. But when $N$ is raised to $600$, the proposed algorithm starts to overtake, and their gap becomes larger when $N$ is further increased. Again, not requiring CSI is a distinct advantage of the proposed algorithm as compared to these benchmarks.
	
\begin{figure}[t]
    \centering
    \includegraphics[width=9cm]{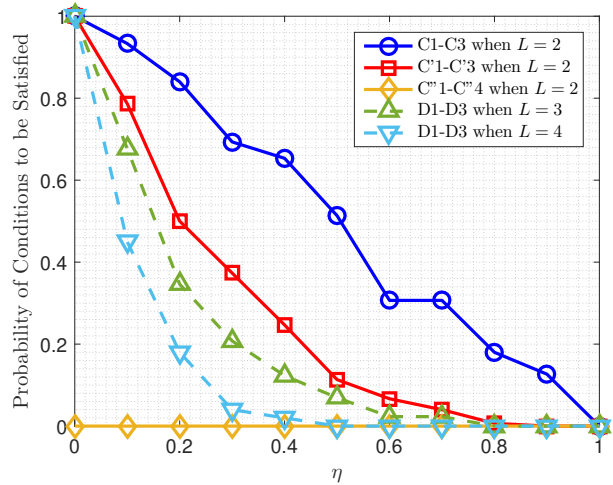}
    \caption{Probability of the conditions for achieving an SNR boost of $\Theta(N^{2L})$ to be satisfied, where $\eta$ is the probability of each channel being LoS.}
    \label{fig:prob_condition}
\end{figure}

Finally, we evaluate how likely the proposed conditions C1-C3 and D1-D3 are satisfied when the IRSs are randomly placed and the channels are randomly generated. We also consider the existing conditions C'1--C'3 in \cite{Han_double_IRS_beamforming_power_scaling} and C''1--C''4 in \cite{Han2022} (but they only account for the double-IRS case). As shown in Fig. \ref{fig:routing_n=600}, the transmitter and the receiver are located at $(5,5)$ and $(95,95)$, respectively. The coordinate vector $(x_\ell,y_\ell)$ of each IRS $\ell$ is uniformly distributed in $[5+\frac{90(\ell-1)}{L},5+\frac{90\ell}{L}]\times[5+\frac{90(\ell-1)}{L},5+\frac{90\ell}{L}]$. Following the indoor field test in Fig. \ref{fig:indoor_layout}, we assume that the IRSs are well placed so that the channels between two adjacent IRSs as well as the transmitter-to-the-first-IRS channel and the last-IRS-to-receiver channel are all LoS. For the rest channels, assume that each is LoS with probability $\eta$. Following 
\cite{Han_double_IRS_beamforming_power_scaling,Han2022}, we assume that the channel equals zero whenever it is NLoS. Let $N=100$ and let each $K_\ell=2L$. In particular, we claim that condition C''4 in \cite{Han2022} is satisfied if the approximation gap is less than $10\%$ of the norm of the original array response.

Fig. \ref{fig:prob_condition} shows how likely the conditions C1--C3, D1--D3, C'1--C'3 and C''1--C''4 are satisfied over 200 tests when the IRSs are randomly placed in the network and the channels are randomly generated; different values of the parameter $\eta$ are considered here. This figure shows that our conditions C1--C3 are much more likely to be satisfied than the existing conditions in the literature. The conditions D1--D3 are less likely to be satisfied since they account for more than 2 IRSs in this case. In general, it is increasingly difficult to satisfy the conditions D1--D3 as more and more IRSs are deployed. In particular, observe that the conditions C''1--C''4 in \cite{Han2022} are extremely difficult to meet; the probability of satisfying C''1--C''4 is almost zero in our case. The reason is that C''1--C''4 can be roughly interpreted as requiring IRS 1 (resp. IRS 2) to be very close to the transmitter (resp. receiver) but such conditions are rarely met in our random network. Also, the conditions C'1--C'3 are less likely to be satisfied than C1--C3 because they require many channels to be zero.

\section{Conclusion}

\label{sec:conclusion}

This work proposes a statistical approach to the multi-IRS beamforming problem in the absence of channel information. For a general $L$-IRS-assisted wireless transmission, we show that the proposed blind beamforming algorithm guarantees an SNR boost of $\Theta(N^{2L})$---which is the highest possible SNR boost obtained from $L$ IRSs, under some certain conditions. This blind beamforming strategy has two major advantages over the existing methods. First, it does not entail any channel estimation and yet can yield provable performance. Second, its optimality condition is far less strict than the existing one in \cite{multi_IRS_Mei}, e.g., those short reflected channels need not be zero for blind beamforming to reach the $\Theta(N^{2L})$ boost. Remarkably, as shown in the real-world experiments at $2.6$ GHz, blind beamforming for multiple IRSs increases SNR by over $17$ dB in the hallway of an office building, and by over $14$ dB near an open caf\'{e}. Moreover, simulations show that much higher gain can be reaped by blind beamforming when the IRSs become larger in size or when more IRSs are deployed.

\section*{Appendix A: Proof of Lemma \ref{lemma gamma+pi/K}}

We establish Lemma \ref{lemma gamma+pi/K} by induction. When $\ell=1$, we have
\begin{align}
\frac{\big|\!\sum_{(n_1,\ldots,n_L)\in\mathcal A^{(1)}_{n_1}}h_{n_1,\ldots,n_L}\big|}{\big|\!\sum_{(n_1,\ldots,n_L)\in\mathcal E^{(1)}_{n_1}}h_{n_1,\ldots,n_L}\!\big|}
&=\frac{\big|\sum_{(n_1,\ldots,n_L)\in\mathcal A^{(1)}_{n_1}}h_{n_1,\ldots,n_L}\big|}{\prod^L_{\ell=2}|\sum^N_{n_\ell=1}u^{(\ell)}_{n_\ell}|\cdot |u^{(1)}_{n_1}|}\notag\\
&\le\frac{\sum_{(n_1,\ldots,n_L)\in\mathcal A^{(1)}_{n_1}}\big|h_{n_1,\ldots,n_L}\big|}{\prod^L_{\ell=2}|\sum^N_{n_\ell=1}u^{(\ell)}_{n_\ell}|\cdot |u^{(1)}_{n_1}|}\notag\\
&\le\sin\gamma,
\label{vec:inequality}
\end{align}
where the last inequality follows by the condition D3 stated in Theorem \ref{Proposition:boost}. Now, with $\ba$ and $\bb$  in Fig. \ref{fig:explanation L=2} redefined as $\sum_{(n_1,\ldots,n_L)\in\mathcal A^{(1)}_{n_1}}h_{n_1,\ldots,n_L}$ and $\sum_{(n_1,\ldots,n_L)\in\mathcal E^{(1)}_{n_1}}h_{n_1,\ldots,n_L}$ respectively, we have
\begin{multline}
\label{appendix:gamma}
\left|
\angle\left(\sum_{(n_1,\ldots,n_L)\in\mathcal A^{(1)}_{n_1}}h_{n_1,\ldots,n_L}
 +\sum_{(n_1,\ldots,n_L)\in\mathcal E^{(1)}_{n_1}}h_{n_1,\ldots,n_L}\right)\right.\\
 -\left.\angle\left(\sum_{(n_1,\ldots,n_L)\in\mathcal E^{(1)}_{n_1}}h_{n_1,\ldots,n_L}\right)\right|\le\gamma,
\end{multline}
which can be recognized as
\begin{equation}
    |\hat\theta^\star_{n_1}-\theta^\star_{n_1}|\le \gamma.
\end{equation}
Combining the above inequality with \eqref{eqn theta' star} yields 
\begin{align}
|\hat\theta^\star_{n_1}-\theta'_{n_1}|\le \gamma+\frac{\pi}{K_1}.
\label{appendix:gamma+}
\end{align}
Thus, Lemma \ref{lemma gamma+pi/K} is verified for $\ell=1$.

Assuming that Lemma \ref{lemma gamma+pi/K} holds for all $\ell<i$, we now proceed to the case of $\ell=i$. It can be shown that
\begin{align} 
&\frac{\big|\sum_{(n_1,\ldots,n_L)\in\mathcal A^{(i)}_{n_i}}h_{n_1,\ldots,n_L}e^{j\sum_{s=1}^{i-1}\theta'_{n_{s}}}\big|}{\big|\sum_{(n_1,\ldots,n_L)\in\mathcal E^{(i)}_{n_i}}h_{n_1,\ldots,n_L}e^{j\sum_{s=1}^{i-1}\theta'_{n_{s}}}\big|}\notag\\
&\overset{(a)}{=}\frac{\big|\sum_{(n_1,\ldots,n_L)\in\mathcal A^{(i)}_{n_i}}h_{n_1,\ldots,n_L}e^{j\sum_{s=1}^{i-1}\theta'_{n_{s}}}\big|}{|\sum_{(n_1,\ldots,n_L)\in\mathcal E^{(i)}_{n_i}}\prod_{s<i}u^{(s)}_{n_s}e^{j\theta'_{n_s}}\cdot \prod_{s>i}u^{(s)}_{n_s}\cdot u^{(i)}_{n_i}|}\notag\\
&\le\frac{\sum_{(n_1,\ldots,n_L)\in\mathcal A^{(i)}_{n_i}}\big|h_{n_1,\ldots,n_L}\big|}{|\sum_{(n_1,\ldots,n_L)\in\mathcal E^{(i)}_{n_i}}\prod_{s<i}u^{(s)}_{n_s}e^{j\theta'_{n_s}}\cdot \prod_{s>i}u^{(s)}_{n_s}\cdot u^{(i)}_{n_i}|}\notag\\
&\overset{(b)}{\le}\frac{\sum_{(n_1,\ldots,n_L)\in\mathcal A^{(i)}_{n_i}}\big|h_{n_1,\ldots,n_L}\big|}{|u^{(i)}_{n_i}|\cdot \prod\limits_{s>i}\!\big|\!\sum^N_{n_s=1}u^{(s)}_{n_s}\big|\cdot\prod\limits_{s<i}\!\big[\!\sum^N_{n_s=1}|u^{(s)}_{n_s}|\cos(\gamma+\frac{\pi}{K_s})\big]}\notag\\
&\overset{(c)}{\le} \sin\gamma,
\end{align}
where step $(a)$ follows by the condition D1 in Theorem \ref{Proposition:boost},  step $(b)$ follows since \eqref{theta_ell:inequality} is assumed to hold for all $\ell<i$, and step $(c)$ follows by the condition D3 in Theorem \ref{Proposition:boost}. Repeating the former steps \eqref{appendix:gamma}--\eqref{appendix:gamma+}, we arrive at
\begin{align}
|\hat\theta^\star_{n_i}-\theta'_{n_i}|\le \gamma+\frac{\pi}{K_i}.
\end{align}
The proof is then completed.

\bibliographystyle{IEEEtran}     
\bibliography{IEEEabrv,strings} 

\begin{IEEEbiographynophoto}{Fan Xu}(Member, IEEE) received the B.S. degree in physics and the Ph.D. degree in information and communication engineering from Shanghai Jiao Tong University, Shanghai,
China, in 2016 and 2022, respectively. He received Huawei Scholarship in 2018 and was the outstanding graduate of Shanghai Jiao Tong University in 2022. 

Since 2022, he joined Peng Cheng Laboratory, Shenzhen, China, as a post-doctor. His research interests include coded caching, distributed computing, intelligent reflecting surface, signal processing and optimization of 5G and beyond networks.
\end{IEEEbiographynophoto}

\begin{IEEEbiographynophoto}{Jiawei Yao}(Student Member, IEEE) 
received the B.E. degree in telecommunication engineering from Zhengzhou University, in 2021. He is currently working toward the M.Phil. degree with the School of Science and Engineering, The Chinese University of Hong Kong, Shenzhen. His research interests include intelligent reflecting surface and integrated sensing and communication.
\end{IEEEbiographynophoto}

\begin{IEEEbiographynophoto}
{Wenhai Lai}(Student Member, IEEE) 
received the B.E. degree in information engineering from Beijing University of Posts and Telecommunications, in 2021. He is currently working toward the Ph.D. degree with the School of Science and Engineering, The Chinese University of Hong Kong, Shenzhen, China. His research interests include intelligent reflecting surface and reinforcement learning.
\end{IEEEbiographynophoto}

\begin{IEEEbiographynophoto}
{Kaiming Shen}(Member, IEEE) received the B.Eng. degree in information security and the B.Sc. degree in mathematics from Shanghai Jiao Tong University, China in 2011, and then the M.A.Sc. degree in electrical and computer engineering from the University of Toronto, Canada in 2013. After working at a tech startup in Ottawa for one year, he returned to the University of Toronto and received the Ph.D. degree in electrical and computer engineering in early 2020. Since 2020, Dr. Shen has been with the School of Science and Engineering at The Chinese University of Hong Kong (Shenzhen), China as a tenure-track assistant professor. His research interests include optimization, wireless communications, information theory, and machine learning.

Dr. Shen received the IEEE Signal Processing Society Young Author Best Paper Award in 2021 for his work on fractional programming for communication systems. Dr. Shen serves as an Editor of IEEE Transactions on Wireless Communications.
\end{IEEEbiographynophoto}

\begin{IEEEbiographynophoto}
{Xin Li}
graduated from Xidian University and joined Huawei in 2008. He has rich experience in wireless channel modeling and wireless network performance modeling and optimization. Currently, he is a technical expert in Huawei's experience lab, focusing on future-oriented network technology research, including new technologies such as Intelligent Reflection Surface and Intelligent Transmission Surface, and their application in network structure optimization.
\end{IEEEbiographynophoto}

\begin{IEEEbiographynophoto}
{Xin Chen}
graduated from the Radio Engineering Department of Southeast University and joined Huawei in 2000. He has 20 years of R\&D experience in the wireless communications field. He has served as senior algorithm engineer, solution architect, and technology development director successively. He has rich experience and achievements in network planning, optimization, and operation and maintenance of mobile communication networks.

Currently, he is the director of the Algorithm Dept of the Service and Software R\&D domain of Huawei Carrier BG. He is responsible for the research of key technologies for digitalization and intelligence in the telecom field, including intelligent network optimization, autonomous network architecture and O\&M, analysis algorithms of telecom big data, and next-generation computing architecture in the telecom field.
\end{IEEEbiographynophoto}

\begin{IEEEbiographynophoto}
{Zhi-Quan (Tom) Luo} (Fellow, IEEE) is the Vice President (Academic) of The Chinese University of Hong Kong, Shenzhen where he has been a professor since 2014. He is concurrently the Director of Shenzhen Research Institute of Big Data. 

Professor Luo received his Ph.D. in Operations Research from MIT in 1989 and his B.S. degree in Mathematics in 1984 from Peking University, China. His research interests lie in the area of optimization, big data, signal processing and digital communication, ranging from theory, algorithms to design and implementation. He served as the Chair of the IEEE Signal Processing Society Technical Committee on Signal Processing for Communications (SPCOM) and the Editor in Chief for IEEE Transactions on Signal Processing (2012--2014), and was an Associate Editor for many internationally recognized journals. 

Professor Luo is a Fellow of the Institute of Electrical and Electronics Engineers (IEEE) and the Society for Industrial and Applied Mathematics (SIAM). He received the 2010 Farkas Prize from the INFORMS Optimization Society, and the 2018 Paul Y. Tseng Memorial Lectureship from the Mathematical Optimization Society. He also received three Best Paper Awards in 2004, 2009 and 2011, a Best Magazine Paper Award in 2015, all from the IEEE Signal Processing Society, and a 2011 Best Paper Award from the EURASIP. In 2014, he was elected to the Royal Society of Canada. Professor Luo was elected to the Chinese Academy of Engineering (as a foreign member) in 2021, and was awarded the Wang Xuan Applied Mathematics Prize in 2022 by the China Society of Industrial and Applied Mathematics.
\end{IEEEbiographynophoto}

\end{document}